\documentclass[final,5p,times,twocolumn,authoryear]{elsarticle}

\journal{}

\newcommand{\nn}{\nonumber} % omit numbering of equation
\newcommand{\ie}{{i.e.},~}
\newcommand{\eg}{{e.g.}~}

\let\cite\citep % redefine cite to behave like citep

\newcommand{\paperTitle}{Spectral Dynamic Causal Modelling: A Didactic Introduction and its Relationship with Functional Connectivity}

\usepackage{amsmath,amssymb,bm,siunitx,mathtools}
\usepackage{graphicx}% Include figure files
\graphicspath{{images/}}
\usepackage{microtype}
\usepackage{tikz}

\usepackage{hyperref} % add hypertext capabilities
\usepackage{color}\definecolor{darkblue}{rgb}{0,0,0.5} % cusotm colors
\hypersetup{unicode=false, pdftoolbar=true, pdfmenubar=true, pdffitwindow=false, pdfstartview={FitH}, pdfcreator={}, pdfproducer={}, pdfkeywords={} {} {}, pdfnewwindow=true, colorlinks=true, linkcolor=red, citecolor=darkblue, filecolor=magenta, urlcolor=darkblue}

\usepackage[capitalise]{cleveref} % automatically figures out if ref is a figure, section, theorem, etc
\Crefname{appendix}{}{}
\usepackage{lineno}
\DeclareMathOperator{\E}{\mathbb{E}} % expectation
\newcommand*\diff{\mathop{}\!\mathrm{d}} %differential for dx in integrals

\begin{document}

\begin{frontmatter}

%% Title, authors and addresses

%% use the tnoteref command within \title for footnotes;
%% use the tnotetext command for theassociated footnote;
%% use the fnref command within \author or \affiliation for footnotes;
%% use the fntext command for theassociated footnote;
%% use the corref command within \author for corresponding author footnotes;
%% use the cortext command for theassociated footnote;
%% use the ead command for the email address,
%% and the form \ead[url] for the home page:
%% \title{Title\tnoteref{label1}}
%% \tnotetext[label1]{}
%% \author{Name\corref{cor1}\fnref{label2}}
%% \ead{email address}
%% \ead[url]{home page}
%% \fntext[label2]{}
%% \cortext[cor1]{}
%% \affiliation{organization={},
%%            addressline={}, 
%%            city={},
%%            postcode={}, 
%%            state={},
%%            country={}}
%% \fntext[label3]{}

\title{\paperTitle}

\author[Turner]{Leonardo Novelli}
\ead{leonardo.novelli@monash.edu}

\author[Wellcome]{Karl Friston}

\author[Turner,Wellcome,CIFAR]{Adeel Razi}

\affiliation[Turner]{
    organization={Turner Institute for Brain and Mental Health, School of Psychological Sciences and Monash Biomedical Imaging, Monash University},%Department and Organization
    country={Australia}}
            
\affiliation[Wellcome]{
    organization={Wellcome Centre for Human Neuroimaging, University College London},
    country={United Kingdom}}

\affiliation[CIFAR]{
    organization={CIFAR Azrieli Global Scholars Program},
    city={Toronto},
    country={Canada}}

\begin{abstract}
We present a didactic introduction to spectral Dynamic Causal Modelling (DCM), a Bayesian state-space modelling approach used to infer effective connectivity from non-invasive neuroimaging data.
Spectral DCM is currently the most widely applied DCM variant for resting-state functional MRI analysis.
Our aim is to explain its technical foundations to an audience with limited expertise in state-space modelling and spectral data analysis.
Particular attention will be paid to cross-spectral density, which is the most distinctive feature of spectral DCM and is closely related to functional connectivity, as measured by (zero-lag) Pearson correlations.
In fact, the model parameters estimated by spectral DCM are those that best reproduce the cross-correlations between all measurements---at all time lags---including the zero-lag correlations that are usually interpreted as functional connectivity.
We derive the functional connectivity matrix from the model equations and show how changing a single effective connectivity parameter can affect all pairwise correlations.
To complicate matters, the pairs of brain regions showing the largest changes in functional connectivity do not necessarily coincide with those presenting the largest changes in effective connectivity.
We discuss the implications and conclude with a comprehensive summary of the assumptions and limitations of spectral DCM.
\end{abstract}

%%%Graphical abstract
%\begin{graphicalabstract}
%\includegraphics{grabs}
%\end{graphicalabstract}

%%%Research highlights
%\begin{highlights}
%\item Research highlight 1
%\item Research highlight 2
%\end{highlights}

%\begin{keyword}
%%% keywords here, in the form: keyword \sep keyword
%keyword one \sep keyword two
%%% PACS codes here, in the form: \PACS code \sep code
%\PACS 0000 \sep 1111
%%% MSC codes here, in the form: \MSC code \sep code
%%% or \MSC[2008] code \sep code (2000 is the default)
%\MSC 0000 \sep 1111
%\end{keyword}

\end{frontmatter}

%\linenumbers

%% For citations use: 
%%       \citet{<label>} ==> Jones et al. (2015)
%%       \citep{<label>} ==> (Jones et al., 2015)

\section{Introduction}
Dynamic causal modelling (DCM) refers to the Bayesian fitting of state-space models to explain observed physiological signals in terms of hidden neuronal activity and connectivity
\cite{friston2003DynamicCausalModelling,triantafyllopoulos2021BayesianInferenceState}.
The distinction between observed and hidden variables is particularly relevant in neuroscience because the signals recorded via non-invasive neuroimaging are not a direct measurement of neuronal states or connectivity.
In fact, using observed recordings to infer unobserved neural interactions is the main purpose of DCM and the reason for its widespread adoption.
It is also a distinctive feature that sets it apart from functional connectivity analysis, which simply characterises statistical dependencies in observed time series.
Over the last 20 years, the versatility offered by state-space models has seen DCM applications in most neuroimaging modalities \cite{friston2003DynamicCausalModelling,kiebel2008DynamicCausalModelling, moran2009DynamicCausalModels,friston2014DCMRestingState,tak2015DynamicCausalModelling,jung2019DynamicCausalModeling,friston2019DynamicCausalModelling,frassle2021RegressionDynamicCausal}, along with recent applications to epidemiology \cite{friston2022DynamicCausalModelling} and beyond \cite{bach2010DynamicCausalModelling}.
Navigating the vast and technical DCM literature, however, is by no means a trivial task---especially to the novice learner.
Happily, there are excellent introductory resources on individual- and group-level analysis using deterministic versions of DCM, which are designed for neuroimaging experiments involving behavioural tasks \cite{stephan2004RoleGeneralSystem,stephan2010TenSimpleRules,zeidman2019GuideGroupEffectivea,zeidman2019GuideGroupEffective}.
A recent primer on variational Laplace explains how Bayesian inference is performed in DCM \cite{zeidman2022PrimerVariationalLaplace} using the SPM software (\url{http://www.fil.ion.ucl.ac.uk/spm}). 
However, there is a lack of introductory material on DCM for resting-state data analysis, despite the remarkable growth of the resting-state paradigm and the widespread uptake  of these methods.
Here, we fill this gap with a didactic introduction to spectral DCM that aims to explain its technical aspects \cite{friston2014DCMRestingState,razi2015ConstructValidationDCM}.

\begin{figure}
    \centering\includegraphics[width=0.5\textwidth]{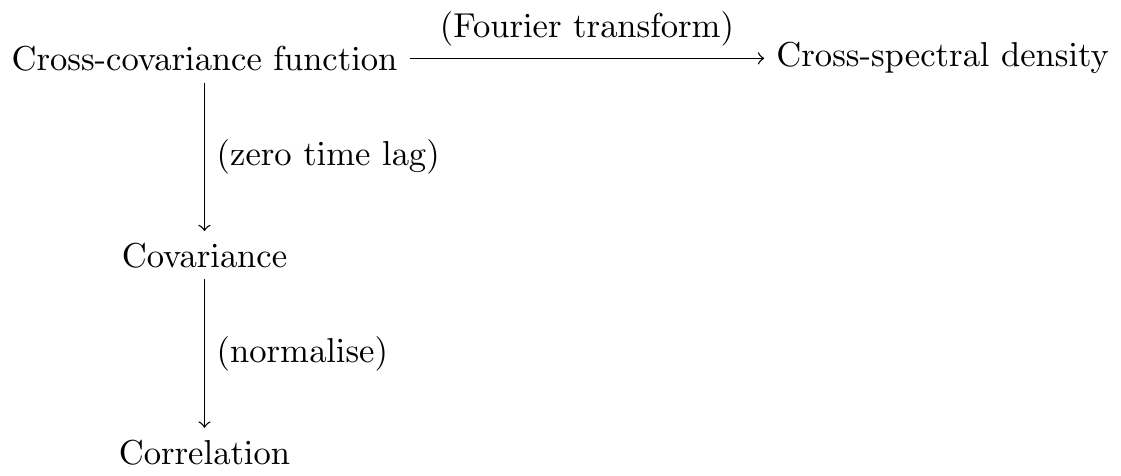}
    \caption{ \label{fig:arrow_diagram_covariance_CSD}
        The distinctive feature of spectral Dynamic Causal Modelling (DCM) is the focus on modelling the measured cross-spectral density (top right), which is a second-order summary statistic of the time-series data.
        This is closely related to Pearson's correlation (bottom left), another second-order statistic and the most widely used measure of functional connectivity in neuroimaging.
        In fact, both can be derived from the cross-covariance function (top left).
        The cross-spectral density is obtained directly via the Fourier transform (horizontal arrow).
        As such, it is the equivalent representation of the cross-covariance function in the frequency domain instead of the time domain.
        The correlation is obtained in two steps (vertical arrows).
        We first compute the covariance as a special case of the cross-covariance function between two time series, by setting the time lag between them to zero.
        Finally, we normalise the covariance such that its values are restricted to the $[-1,1]$ interval to obtain the zero-lag correlation.
        For the mathematical relationships among these quantities, we refer readers to \cite[Fig. 1]{friston2014DCMRestingState}.
    }
\end{figure}

What distinguishes spectral DCM from other DCM versions, and when should we choose it?
Firstly, spectral DCM employs random differential equations instead of deterministic ones.
These are used to model spontaneous endogenous fluctuations in neuronal activity, enabling resting-state analysis in the absence of experimental inputs.
But its distinctive feature is the focus on modelling the measured cross-spectral density, which is a second-order summary statistic of the time-series data.
This is closely related to Pearson's correlation, another second-order statistic and the most widely used measure of functional connectivity in neuroimaging (see diagram in \cref{fig:arrow_diagram_covariance_CSD}; for mathematical relationships, also see \cite[Fig. 1]{friston2014DCMRestingState}).
In fact, the correlation is obtained by normalising the covariance such that its values are restricted to the $[-1,1]$ interval.
In turn, the covariance is a special case of the cross-covariance function between two time series, when there is no time lag between them.
Finally, the Fourier transform of the cross-covariance function gives the cross-spectral density (under stationarity assumptions).
In other words, the cross-spectral density is the equivalent representation of the cross-covariance function in the frequency domain instead of the time domain---an important relationship that we will unpack later in this article. 

Spectral DCM fits the parameters of a linear, continuous-time model to the observed cross-spectral density.
The estimated parameters are those that best reproduce the cross-correlations between all variables, at all time lags.
In particular, the estimated effective connectivity also reproduces the zero-lag correlations between the observed time series---the most common measure of functional connectivity in the literature.
This would be appealing to researchers who are interested in both effective and functional connectivity.
The nuanced relationship between effective and functional connectivity is explored in \cref{sec:EC_vs_FC}.
Prior to that, in \cref{sec:generative_model}, we introduce and explain the various components of the generative model, \ie the model that generates the cross-spectral density given a set of parameters.
These basic building blocks are used routinely in signal processing and control theory and are often presented only briefly in the DCM literature.
Here, we adopt an inclusive and slower pace for those who are not familiar with state-space models and spectral data analysis.
That said, we count on the reader to fill in the gaps and look up concepts such as the Fourier transform \cite{oppenheim1997SignalsSystems,smith2002ContinuousSignalProcessing} or convolution \cite{smith2002ContinuousSignalProcessing}, if needed.
Moving from theory to practice, a step-by-step guide to running spectral DCM on a real resting-state functional MRI (fMRI) dataset is provided in Chapter 38 of the SPM12 manual \cite{johnashburner2020SPM12Manual}.

A final reason for choosing spectral DCM is its computational advantage compared to stochastic DCM \cite{li2011GeneralisedFilteringStochastic}.
It is important to note that the lower computational complexity and the resulting increase in speed rely on the assumption that the the statistics of endogenous neuronal fluctuations are conserved over the experimental time window, making spectral DCM suitable for resting-state neuroimaging experiments.
Experimental inputs can also be included via an additional term in the model, although applications to task experiments are infrequent in the literature.
Introducing even stronger assumptions leads to even faster schemes, such as \emph{regression DCM}, which can analyse hundreds of brain regions in minutes \cite{frassle2021RegressionDynamicCausal}.
However, this method forgoes the strict separation between hidden and observed variables that is typical of state-space modelling and that we have used to define DCM herein.
As the name suggests, regression DCM is more akin to the Bayesian fitting of a multivariate autoregressive model in the frequency domain.

The key assumptions made in spectral DCM are summarised in \cref{sec:assumptions} with a discussion of the ensuing limitations.

\section{Building the generative model, one element at a time}
\label{sec:generative_model}
The signals recorded via non-invasive neuroimaging are not a direct measurement of neuronal activity.
In the case of fMRI, the observed blood-oxygen-level-dependent (BOLD) signal captures changes in blood oxygenation that indirectly reflect neuronal activity.
For this simple reason, spectral DCM models the neuronal and the observed variables separately (denoted by $x$ and $y$, respectively).
Such a distinction represents both the main strength and the challenge of the DCM framework.

\subsection{Neuronal model}
\label{sec:neuronal_model}
In spectral DCM, the neuronal model is defined by the linear random differential equation
\begin{equation} \label{eq:neuronal_model_full}
    \bm{\dot{x}}(t) = A \bm{x}(t) + \bm{v}(t),
\end{equation}
where $\bm{x}(t)$ is the $n$-dimensional \emph{state vector}
\begin{equation}
    \bm{x}(t) =
    \begin{bmatrix}
        x_1(t) \\
        \vdots \\
        x_N(t) 
    \end{bmatrix}  \nn
\end{equation}
whose $N$ scalar components represent different brain regions.
These are called \emph{state variables} in the state-space modelling literature \cite{durbin2012TimeSeriesAnalysis,williamsii2007StateSpaceFundamentals}.
The time derivative of the state vector is denoted as $\bm{\dot{x}}(t)$, where differentiation with respect to time is performed component-wise, \ie $\bm{\dot{x}}(t) = [ \dot{x}_1(t), \ldots, \dot{x}_N(t) ]^\intercal$.
The activity of the system is sustained by stochastic, non-Markovian, endogenous fluctuations denoted as $\bm{v}(t)$, which we will consider in \cref{sec:endogenous_fluctuations}.
Let us first turn our attention to the $A$ matrix, which defines and parameterises the \emph{effective connectivity}.

\subsection{Effective connectivity}
\label{sec:effective_connectivity}
The effective connectivity quantifies the directed effect of one brain region on another.
To better understand its meaning, consider a simple deterministic system with two brain regions and no stochastic components.
The matrix-vector notation in \cref{eq:neuronal_model_full} can be unpacked into two scalar components
\begin{subequations} \label{eq:neuronal_eq_two_variables}
\begin{align}
        \dot{x}_1(t) &= a_{11} x_1(t) + a_{12} x_2(t)  \label{eq:neuronal_first_var} \\
        \dot{x}_2(t) &= a_{21} x_1(t) + a_{22} x_2(t),
\end{align}
\end{subequations}
where $a_{jk}$ corresponds to the element in the $j$-th row and $k$-th column of $A$.
More explicitly, in this example, we have
\begin{equation}
  A =
  \left[ {\begin{array}{cc}
    a_{11} & a_{12} \\
    a_{21} & a_{22} \\
  \end{array} } \right].
\end{equation}
Let's initially assume that $x_1$ is inactive at time $t_1$ and set $x_1(t_1)=0$ in \cref{eq:neuronal_first_var}.
We get
\begin{equation}
    \dot{x}_1(t_1) = a_{12} x_2(t_1),
\end{equation}
stating that the instantaneous change in $x_1$ is proportional to the input from $x_2$.
The effective connectivity $a_{12}$ is simply the coefficient that determines the \emph{rate} of such change.
Therefore, in DCM, the effective connectivity $a_{jk}$ quantifies the instantaneous response rate of $x_{j}$ caused by a change in $x_{k}$, in the ideal case where all other variables were kept fixed or set to zero.\footnote{Readers who are familiar with multivariate calculus would recognise this as a partial derivative and the $A$ matrix as a Jacobian.}
Being a rate, effective connectivity is always measured in Hz (change per second).
\begin{figure}
    \centering\includegraphics[width=0.5\textwidth]{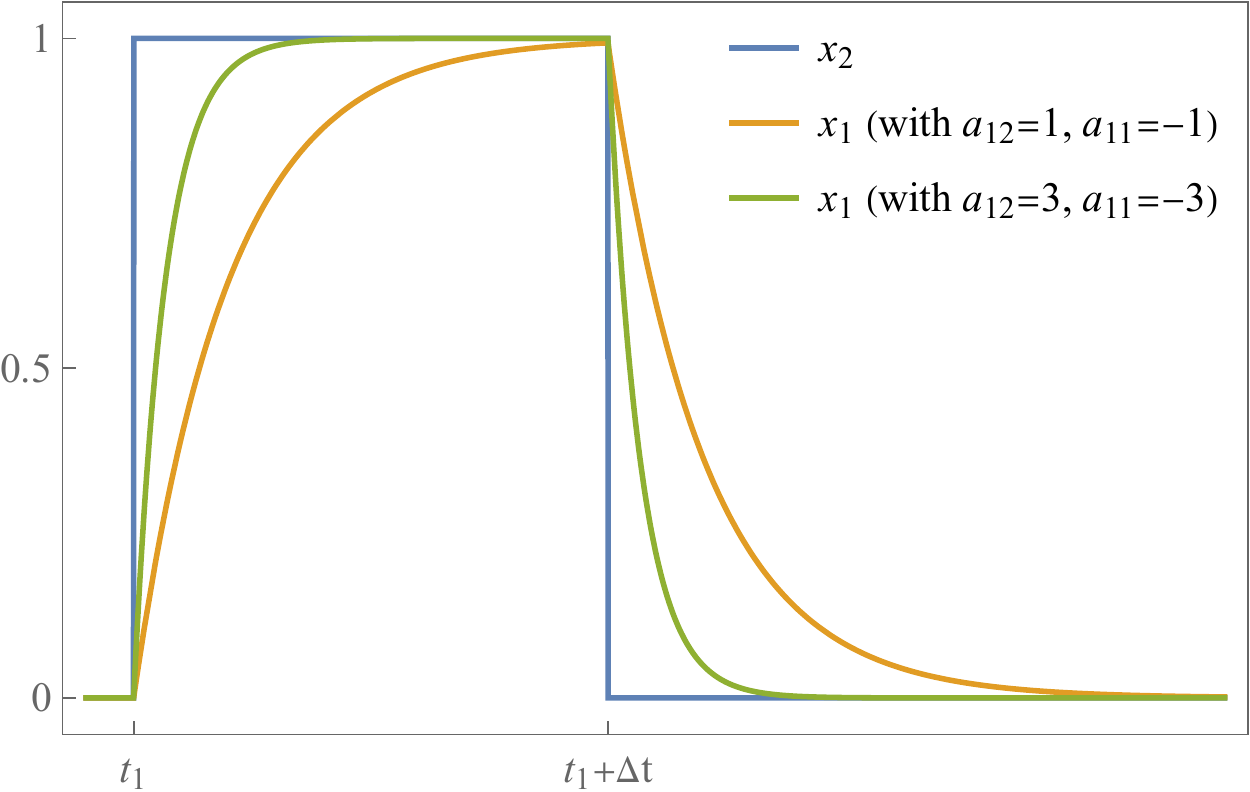}
    \caption{\label{fig:linear_response_2D}
        The role of the effective connectivity parameters in a deterministic linear system with two variables.
        The parameter $a_{12}$ determines the instantaneous rate of response of $x_1$ to an input from $x_2$.
        In this example, the input is constant with duration $\Delta t$.
        The initial slope of the response is steeper for higher values of $a_{12}$.
        Once the input from $x_2$ ceases, the activity of $x_1$ decays exponentially in the absence of other inputs.
        The decay rate is determined by the self-connection $a_{11}$, with larger negative values resulting in faster decay (shorter memory).
    }
\end{figure}
\cref{fig:linear_response_2D} shows the impact of $a_{12}$ on the response of $x_1$ to a constant input from $x_2$ with duration $\Delta t$.
Note that the effective connectivity determines the initial slope of the curve, which is steeper for high values of $a_{12}$.
However, once the input from $x_2$ ceases, the magnitude and duration of the response in $x_1$ no longer depend on $a_{12}$; instead, they only depend on the self-connection $a_{11}$.
That is, after the time interval $\Delta t$, we have $x_2(t)=0$ and \cref{eq:neuronal_first_var} becomes
\begin{equation} \label{eq:self-connection}
    \dot{x}_1(t) = a_{11} x_1(t),
\end{equation}
which has the simple exponential solution
\begin{equation} \label{eq:exponential_solution}
    x_1(t) = c e^{a_{11} t},
\end{equation}
for each time $t>(t_1+\Delta t)$, where the constant factor $c$ is the value of $x_1$ when the input from $x_2$ ceases, \ie $c=x_1(t_1+\Delta t)$.
It is useful, and biologically-plausible, to impose a negativity constraint on the rate constant of the self-connections (\ie $a_{11}$) to avoid instability and divergence to infinity.
In this example, a negative value of $a_{11}$ guarantees that $x_1(t)$ converge to zero.\footnote{
The reason why DCM studies often report positive values on the diagonal of the effective connectivity matrix $A$ is that the self-connections are transformed using the logarithmic function $\log(-2a)$ by SPM.
This convention is technically motivated by the use of log-normal priors to enforce positivity or negativity on certain parameters; here, to enforce recurrent or self inhibition.
A reported zero value for a self-connection corresponds to $-0.5$~Hz, which is the default prior self-connectivity value in SPM12.
Positive values correspond to slower decay rates in the $(-0.5,0)$ range, while negative values correspond to faster decays ($<-0.5$).
}
In the multivariate case, the stability of a linear dynamical system is guaranteed when all the eigenvalues of the effective connectivity matrix $A$ have a negative real part \cite{izhikevich2006TwoDimensionalSystems}.

\subsection{Power spectral density and cross-spectral density}
\label{sec:PSD_and_CSD}
A time-varying signal $z(t)$ is a function of time.
However, the same function could be represented as a sum of elementary sine waves, each characterised by a single frequency.
This sum is weighted, with some frequencies carrying more weight than others (\ie the sine waves can have different amplitudes).
Every function, $z(t)$, is a unique mix of frequencies, some more pronounced, some less.
This unique profile is called Fourier frequency spectrum.
The time- and frequency-domain representations of a function are two sides of the same coin: they are equally informative but reveal different and complementary aspects of the same data.
The Fourier transform ($\mathcal{F}$) is the mathematical tool that turns the coin over: it converts a function of time into the corresponding function of frequency (while the inverse Fourier transform does the opposite).
If we denote the (angular) frequency by $\omega$, then $\mathcal{F}$ turns the time function $z(t)$ into the frequency function $Z(\omega)$.
Mathematically, this transformation is achieved via the integral
\begin{equation} \label{eq:Fourier}
     Z(\omega) = \mathcal{F}\{ z(t) \} = \int^{\infty }_{-\infty } z(t) e^{-i \omega t} \diff t
\end{equation}
(for time signals, we often only consider positive values of $t$ and compute the integral in the $[0,\infty]$ interval; this is equivalent to setting $z(t)=0$ for all $t<0$).
The resulting $Z(\omega)$ is a function of the frequency $\omega$ and no longer depends on time.
Somewhat ambiguously, the term ``Fourier transform'' is used both to denote the mathematical operation and the resulting function, $Z(\omega)$.
Note that $Z(\omega)$ typically returns complex values, due to the presence of the imaginary unit $i$ in \cref{eq:Fourier}.
Yet, the squared magnitude of a complex number (\eg $|Z(\omega)|^2$) is a real number, defined as the square of its real part plus the square of its complex part.
Therefore, the magnitude of the Fourier transform is a function that only returns real values, which makes it easier to understand and visualise.
This function is the \emph{power spectral density} of the signal.
The simplicity of interpretation comes with of a loss of information.
After computing the squared magnitude to obtain the power spectral density, we cannot go back and recover the original complex-valued Fourier transform (similarly to what happens for real numbers, where the square produces a unique result but the square root has two solutions).
A similar information loss affects all second-order properties of the time series, including the cross-spectral density and the (cross-) correlation, which are two key concepts that we will discuss and connect later.

Until now, we have only considered deterministic signals.
However, spectral DCM is concerned with stochastic (non-deterministic) processes, such as the endogenous fluctuations that we will examine in the next section, which are a proxy for thoughts or mind-wandering-like processes during resting-state brain activity.
A stochastic process is a sequence of random variables.
If $x$ is a stochastic process indexed by time, then $x(t)$ is not a number but a random variable with a given probability distribution (see \cref{app:stochastic_processes} for an illustration).
The simplest example is the white noise process, which follows a Normal distribution at each time point.
According to stochastic calculus, the Fourier transform of white noise is also a stochastic process; however, it is indexed by frequency instead of time.
Specifically, each frequency $\omega$ corresponds to a random variable that follows a Normal distribution with unit variance.
Since the variance is the same for all frequencies, white noise has a flat power spectral density.
The mathematically-versed reader would have noticed that the power spectral density of a stochastic process is also a stochastic process indexed by frequency, obtained as the squared magnitude of the Fourier transform.
Therefore, in the case of stochastic processes, we will consider the expected value of the power spectral density ($\E[|Z(\omega)|^2]$), which is a number, \ie a scalar function of frequency.
Being the expectation of the squared magnitude, the power spectral density can also be understood as the variance of the Fourier transform of a stochastic process, if the latter has zero mean.
We can equivalently express this concept via the equation
\begin{equation}
    \textup{Var}[\mathcal{F}\{ z(t) \}] = \textup{Var}[Z(\omega)]=\E[|Z(\omega)|^2]+|{\underbrace{\E[Z(\omega)]}_{=0}}|^2=\E[|Z(\omega)|^2].
\end{equation}
(Technical note: for simplicity, we assume that the Fourier transform exists. The general definition of the power spectral density involves a limit over the bounds of the Fourier integral \cite{miller2012PowerSpectralDensity}).

From here, we can seamlessly transition to multivariate stochastic processes using the same mathematical tools.
If $\bm{x}(t)$ is a vector with one element per brain region, its Fourier transform $\bm{X}(\omega)$ is also a vector.
This is important because it applies to the stochastic neuronal variable in \cref{eq:neuronal_model_full}.
The multivariate analogue of the power spectral density is the \emph{cross-spectral density}, defined as the covariance matrix
\begin{equation} \label{eq:CSD_neuronal}
    G_x(\omega) = \textup{cov}[\bm{X}(\omega), \bm{X}(\omega)] =  \E[\bm{X}(\omega) \bm{X}(\omega)^\dag],
\end{equation}
where $\dag$ indicates the conjugate transpose of the vector $\bm{X}(\omega)$.
The dot product between the column vector $\bm{X}(\omega)$ and its conjugate transpose is a square matrix.
Specifically, $G_x(\omega)$ is a $N \times N$ matrix whose diagonal elements are the power spectral densities (variances) of individual neuronal variables, representing various brain regions.
These are real positive numbers.
Each off-diagonal element describes the cross-spectral density (covariance) between a different pair of variables.
Unlike the diagonal elements, they generally take complex values. 
%since the imaginary part doesn't cancel out in the product.

Admittedly, the cross-spectral density definition as a covariance in the frequency domain is quite abstract.
A better intuition will develop after exploring the close relationship between cross-spectral density and functional connectivity.
In \cref{sec:CSD_Khinchin}, we'll see how the cross-spectral density is the Fourier transform of the cross-covariance function, which captures both the correlation matrix and its time-lagged extensions.
For now, the power spectral density definition given above is sufficient to understand how endogenous fluctuations are modelled in spectral DCM.

\subsection{Endogenous fluctuations}
\label{sec:endogenous_fluctuations}
Stable deterministic linear systems can only converge to a state of permanent equilibrium or produce an infinite sequence of identical oscillations.
To overcome these limitations and add variability to the neuronal oscillations, we can introduce endogenous (intrinsic) fluctuations in the system, sometimes referred to as \emph{state noise}.
For example, adding the stochastic term $v_1(t)$ to \cref{eq:self-connection} gives:
\begin{equation} \label{eq:self-connection_stochastic}
    \dot{x}_1(t) = a_{11} x_1(t) + v_1(t).
\end{equation}
At each time $t$, the random variable $v_1(t)$ provides an endogenous input to the neuronal variable so that $x_1(t)$ doesn't converge to zero despite the negative self-decay rate $a_{11}$.
This holds true even in the absence of experimental inputs and inputs from other variables, as is the case in \cref{eq:self-connection_stochastic}.
In other words, the neuronal activity is now also modelled as an intrinsically fluctuating signal, \ie a stochastic process.
The addition of a stochastic term to a dynamical system is traditionally used to model noise, often assumed to be white (that is, serially uncorrelated and with a flat spectral density).
Spectral DCM relaxes this assumption and allows the endogenous fluctuations to be temporally correlated, which makes them non-Markovian and smooth. 
%\footnote{Their eigenvalues tend towards zero from below and show critical slowing.}
Specifically, their power spectral density is modelled to follow a power-law decay as a function of the frequency $\omega$:
\begin{equation} \label{eq:endogenous_fluct_spectrum}
    G_{v_j}(\omega) = \alpha_{v_j} \omega^{-\beta_{v_j}}.
\end{equation}
The parameters $\alpha_{v_j}$ and $\beta_{v_j}$ determine the amplitude and the decay rate of the power-law and may differ between neuronal regions ($j=1,\ldots,N$).
Note that the power-law family includes the flat spectrum (white noise) as a special case where $\beta_{v_j}=0$.

The endogenous fluctuations driving one neuronal variable are assumed to be independent of those driving the others.
The result is that the cross-spectral density of the endogenous fluctuations vector $\bm{v}(t)=[v_1(t),\ldots,v_N(t)]^\intercal$ is a diagonal matrix.
More precisely, the Fourier transform of $\bm{v}(t)$, denoted as the vector $\bm{V}(\omega)$, is a multivariate Gaussian random variable with zero mean and diagonal covariance matrix
\begin{equation}
    G_{v}(\omega)=\E[\bm{V}(\omega) \bm{V}(\omega)^\dag]=\textup{diag}[G_{v_j}(\omega)], \text{for } j=1,\ldots,N,
\end{equation}
where each diagonal entry is the power spectral density of an endogenous fluctuation variable.
We will return to this expression when assembling all the elements of the generative model.

\subsection{Observation function}
\label{sec:observation_function}
\textls[-15]{
We motivated the use of state-space models by their ability to distinguish between hidden and observed variables.
The function that relates the two is known as the \textit{observation function}.
Imagine hearing thunder, where the sound (observed variable) is generated by lightning (hidden variable).
The role of the observation function is to describe the intensity and delay of the sound based on the distance from the lightning.
The specific observation function used in fMRI is the hemodynamic response function (HRF), which links the neuronal activity to the observed BOLD signal.
Similarly to the lightning and thunder example, there is a delay between the neuronal activity and the ensuing peak of the BOLD response.
The profile of the response depends on several region-specific biophysical parameters and can be modelled mathematically \cite{stephan2007ComparingHemodynamicModels}.
For simplicity, we will denote the HRF of a brain region $j$ as $h_j(t)$, without explicitly indicating the biophysical parameters.
The BOLD signal $y_j(t)$ is obtained via convolution of the HRF with the neuronal activity:
\begin{equation} \label{eq:HRF_convolution_single}
    y_j(t) = h_j(t) \ast x_j(t) + e_j(t),
\end{equation}
where $j=1,\ldots,N$ and $e_j(t)$ denotes the observation noise.
By analogy with the endogenous fluctuations in \cref{eq:endogenous_fluct_spectrum}, spectral DCM assumes that the power spectral density of the observation noise also follows a power-law decay:
\begin{equation} \label{eq:observation_noise_spectrum}
    G_{e_j}(\omega) = \alpha_{e_j} \omega^{-\beta_{e_j}}.
\end{equation}
The equivalent representation in vector notation is
\begin{align}
    \bm{y}(t)       &= h(t) \ast \bm{x}(t) + \bm{e}(t) \nn \\
    \bm{E}(\omega)  &\sim \mathcal{N}(\textbf{0},\,G_{e}(\omega)),
\end{align}
where $h(t)$ is a diagonal matrix with diagonal elements $h_j(t)$ for $j=1,\ldots,N$ (as before, $N$ is the number of regions).
The noise terms in the vector $\bm{e}(t)=[e_1(t),\ldots,e_N(t)]^\intercal$ are assumed to be independent of each other, \ie the noise in each region is independent of the noise in the other regions.
Thus, the Fourier transform of $\bm{e}(t)$, denoted as $\bm{E}(\omega)$, is a multivariate Gaussian random variable with zero mean and diagonal covariance matrix $\E[\bm{E}(\omega) \bm{E}(\omega)^\dag]=G_{e}(\omega)$, whose diagonal entries are $G_{e_j}(\omega)$, for all $j=1,\ldots,N$. 
When working in the frequency domain, we can implement the haemodynamic response function as a filter---usually suppressing high frequencies---and implementing delays by operating on the imaginary parts of the Fourier coefficients.

\subsection{Putting it all together}
\label{sec:CSD_BOLD}
The full state-space model used in spectral DCM is
\begin{align}\label{eq:generative_model}
    \dot{\bm{x}}(t)      &= A \bm{x}(t) + \bm{v}(t) \nn \\
    \bm{y}(t)       &= h(t) \ast \bm{x}(t) + \bm{e}(t) \nn \\
    \bm{V}(\omega)  &\sim \mathcal{N}(\textbf{0},\,G_{v}(\omega)) \nn \\
    \bm{E}(\omega)  &\sim \mathcal{N}(\textbf{0},\,G_{e}(\omega)),
\end{align}
where each equation has been introduced in its respective section above.
Since DCM uses a Bayesian framework, all the model parameters are equipped with a prior distribution \cite{friston2014DCMRestingState}.
Their posterior distribution is then computed via Bayesian inference using variational Laplace \cite{zeidman2022PrimerVariationalLaplace,friston2007VariationalFreeEnergy}.
In spectral DCM, this goal is achieved by fitting the generative model in \cref{eq:generative_model} to the cross-spectral density of the data\footnote{In SPM12, the cross-spectral density is estimated in a parametric fashion, by fitting a vector autoregressive model to the observed time series and leveraging its properties.
The default model order used for the slow BOLD fluctuations is $8$. Also note that here, for simplicity, we have omitted the additional term describing the sampling error of the cross-spectral density \cite{friston2014DCMRestingState}.}, 
which is defined as the matrix
\begin{align} \label{eq:CSD_BOLD_definition}
    G_y(\omega)     &=  \E[ \mathcal{F}\{ \bm{y}(t) \} \, \mathcal{F}\{ \bm{y}(t) \}^\dag ].
\end{align}
Going ``backwards'' from the observed cross-spectral density to the parameter distribution is only possible once we specify the forward model.
Specifying the forward model means to derive $G_y(\omega)$ as a function of the model parameters.
We can start by invoking the convolution theorem, which states that the Fourier transform of a convolution of two functions is the (dot) product of their Fourier transforms.
In the case of the observed signal $\bm{y}(t)$, we get
\begin{align}
    \mathcal{F}\{ \bm{y}(t) \}  &= \mathcal{F}\{ h(t) \ast \bm{x}(t) + \bm{e}(t) \} \nn \\
                            &= \mathcal{F}\{ h(t) \} \, \mathcal{F}\{ \bm{x}(t) \} + \mathcal{F}\{ \bm{e}(t) \} \nn \\
                            &= H(\omega) \bm{X}(\omega) + \bm{E}(\omega) \label{eq:Fourier_y},
\end{align}
where the terms in capital letters are the Fourier transforms of the corresponding lowercase functions.
Linear control theory gives us a useful expression for $\bm{X}(\omega)$, obtained as the solution to the linear differential equation in \cref{eq:neuronal_model_full} via the Laplace method:
\begin{equation} \label{eq:resolvent_multivariate}
    \bm{X}(\omega) = (i\omega I-A)^{-1} \bm{V}(\omega), 
\end{equation}
where $I$ is the $N$-dimensional identity matrix \cite{pipes1968ApplicationsLaplaceTransforms}.
Plugging \cref{eq:Fourier_y,eq:resolvent_multivariate} into \cref{eq:CSD_BOLD_definition} yields
\begin{align}
    G_y(\omega) =& \E[ H(\omega) \bm{X}(\omega) \bm{X}(\omega)^\dag H(\omega)^\dag ] + \E[\bm{E}(\omega) \bm{E}(\omega)^\dag] \nn \\
                &+ \E[ H(\omega) \bm{X}(\omega) ] \E[ \bm{E}(\omega) ]^\dag + \E[ \bm{E}(\omega) ]  \E[ H(\omega) \bm{X}(\omega) ]^\dag \label{eq:CSD_expectations} \\
                %=& H(\omega) \E[{|\bm{X}(\omega)|}^2] H(\omega)^\dag + \E[{|E(\omega)|}^2] \\
                %=& H(\omega) G_x(\omega) H(\omega)^\dag + G_{e_j}(\omega) \\
                =& \E[ H(\omega) (i\omega I-A)^{-} \bm{V}(\omega) \bm{V}(\omega)^\dag  (-i\omega I-A^\intercal)^{-} H(\omega)^\dag ] \nn \\
                &+ \E[\bm{E}(\omega) \bm{E}(\omega)^\dag] \\
                =& H(\omega) (i\omega I-A)^{-} G_{v}(\omega) (-i\omega I-A^\intercal)^{-} H(\omega)^\dag + G_{e}(\omega) \label{eq:CSD_BOLD_forward}
\end{align}
The last two terms in \cref{eq:CSD_expectations} vanish since the noise term $\bm{E}(\omega)$ is a Gaussian random variable with zero mean, \ie $\E[ \bm{E}(\omega) ]=0$.
In \cref{eq:CSD_BOLD_forward}, we have substituted $\E[\bm{V}(\omega) \bm{V}(\omega)^\dag]=G_{v}(\omega)$ and $\E[\bm{E}(\omega) \bm{E}(\omega)^\dag]=G_{e}(\omega)$, as defined in \cref{sec:endogenous_fluctuations,sec:observation_function}.

We have now fully described the forward model used in spectral DCM.
The predicted cross-spectral density can be computed using the following model parameters:
\begin{enumerate}
    \item{the effective connectivity parameters in the $A$ matrix};
    \item{the power-law parameters (\ie the amplitude and the exponent) describing the spectrum of the endogenous fluctuations and the observation noise (\cref{eq:endogenous_fluct_spectrum,eq:observation_noise_spectrum})};
    \item{the observation function parameters, \eg the biophysical parameters of the BOLD balloon model}.
\end{enumerate}
Crucially, neither the neuronal state variables $\bm{X}(\omega)$ nor the endogenous fluctuations $\bm{V}(\omega)$ appear in \cref{eq:CSD_BOLD_forward}, only the parameters describing their cross-spectral densities.
This parametrisation allows spectral DCM to infer the model parameters listed above without inferring the hidden neuronal states.
Inferring the state variables (neuronal time series) is a computationally harder problem addressed by stochastic DCM \cite{li2011GeneralisedFilteringStochastic}.
}

\subsection{Simulated and empirical cross-spectral density}
\label{sec:simulated_and_empirical_CSD}
\begin{figure}
    \centering\includegraphics[width=0.5\textwidth]{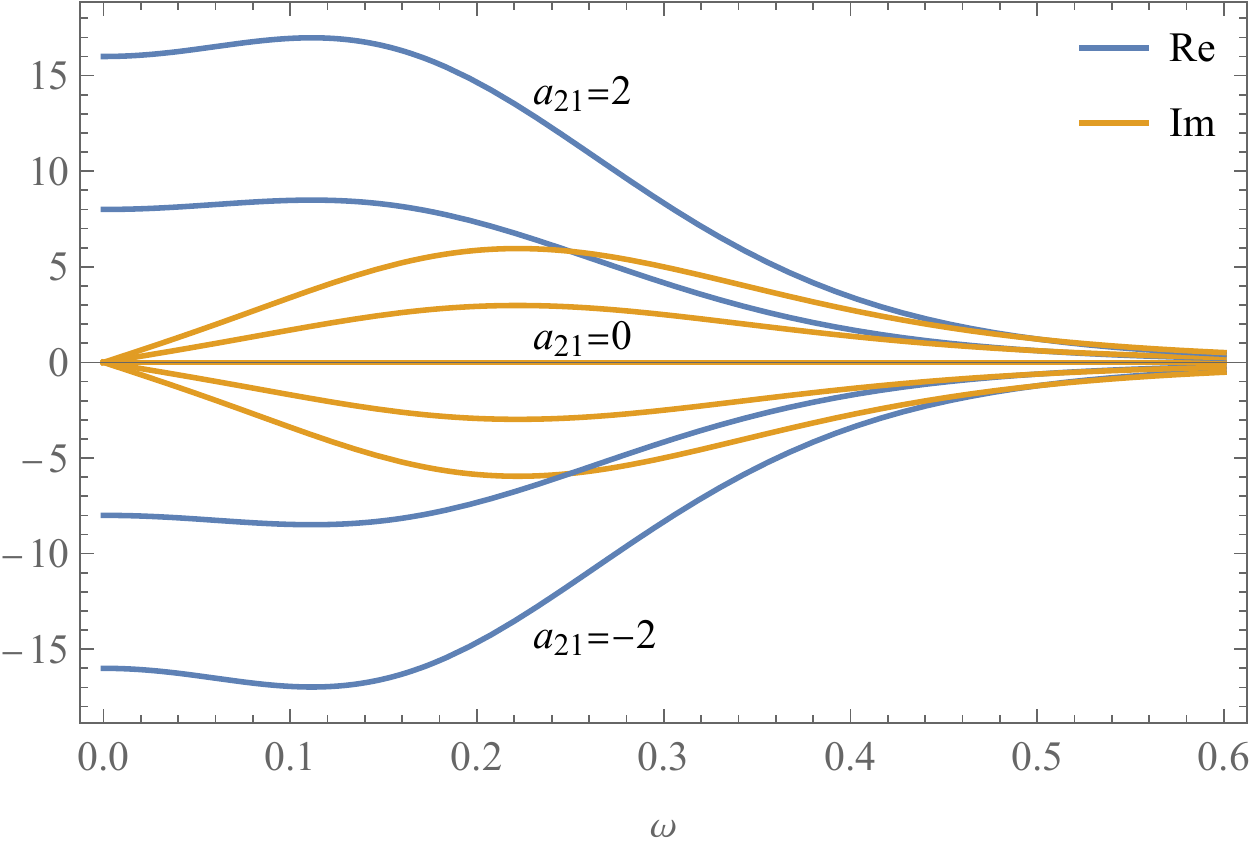}
    \caption{\label{fig:CSD_y12_span}
        Cross-spectral density of the BOLD signal as a function of the effective connectivity between two neuronal state variables ($a_{21}$).
        In this first example, increasingly large and positive values of $a_{21}$ generate increasingly large and positive cross-spectral density amplitudes, while negative values generate negative amplitudes.
        However, this monotonic relationship doesn't hold in more realistic scenarios.
        When $a_{21}=0$, the two neuronal state variables are independent and the cross-spectral density is zero at all frequencies.
    }
\end{figure}

The forward model enables Bayesian inversion and also allows us to understand how different parameters affect the observed cross-spectral density.
\cref{fig:CSD_y12_span} shows how the cross-spectral density $[G_y(\omega)]_{21}$ varies in a system with two hidden neuronal state variables, as a function of their effective connectivity strength.
Specifically, the chosen effective connectivity matrix is
\begin{equation} \label{eq:A_2D_example}
  A =
  \left[ {\begin{array}{cc}
    -\frac{1}{2} & 0 \\
    a_{21} &-\frac{1}{2} \\
  \end{array} } \right].
\end{equation}
The asymmetry in $A$ indicates a directed effect of the first state variable on the second, but not vice-versa.
The strength of the connection is determined by $a_{21}$.
In this first simple example, increasingly large and positive values of $a_{21}$ generate increasingly large and positive cross-spectral density amplitudes.
Similarly, negative values generate negative amplitudes (but we'll soon encounter more complex scenarios that violate this monotonic relationship).
When $a_{21}=0$, the two neuronal state variables are independent of each other and the cross-spectral density is zero at all frequencies (\cref{fig:CSD_y12_span}; also see \cref{app:real_imaginary_CSD} for an explanation of the real and imaginary parts).
Here, the cross-spectral densities of the endogenous fluctuations and of the observation noise are identical to each other and identical for both state variables (so that both diagonal entries are equal):
\begin{equation} \label{eq:power_law_decay_2D_example}
  G_{v}(\omega)=G_{e}(\omega)=
  \left[ {\begin{array}{cc}
    \frac{1}{(1+\omega)^{2}} & 0 \\
    0 &\frac{1}{(1+\omega)^{2}} \\
  \end{array} } \right]
  = (1+\omega)^{-2} \; I
\end{equation}
The observation function is also identical for both state variables.
Spectral DCM employs a HRF model with region-specific biophysical parameters \cite{stephan2007ComparingHemodynamicModels}.
However, for simplicity, the examples in this section are based on the canonical HRF gamma-mixture, whose Fourier spectrum is
\begin{equation} \label{eq:HRF_canonical_Fourier}
    H(\omega)=\frac{6 (i\omega+1)^{10}-1}{5 (i\omega+1)^{16}} \; I.
\end{equation}
High-frequencies are `cut off' since the slow HRF smooths the faster neuronal activity (a typical feature of the BOLD signal).

\begin{figure}[h!]
    \centering\includegraphics[width=0.25\textwidth]{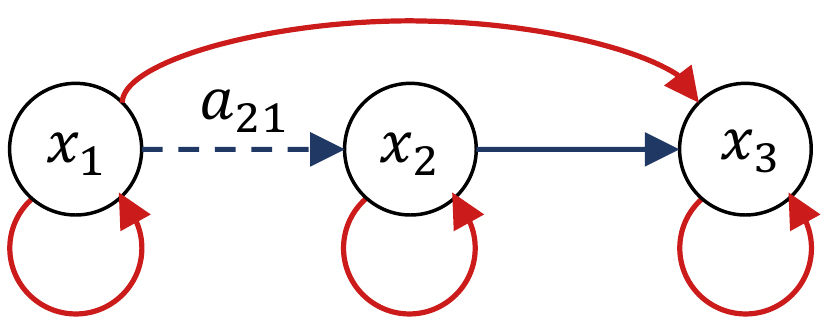}
    \caption{\label{fig:3_variables_network_graph}
        Network graph of the system described by the effective connectivity matrix $A$ in \cref{eq:A_3D_example}.
        The strength of the directed connection from $x_1$ to $x_2$ is set via the parameter $a_{21}$.
        The third neuronal state variable receives an inhibitory influence from $x_1$ and an excitatory influence from $x_2$.
        All three variables have the same negative self-connections represented by the diagonal elements of $A$.
    }
\end{figure}

Let's now extend the system in \cref{eq:A_2D_example} by adding a third neuronal state variable (see \cref{fig:3_variables_network_graph}) and a third row and column to the effective connectivity matrix:
\begin{equation} \label{eq:A_3D_example}
  A =   \left[
        \begin{array}{ccc}
         -\frac{1}{2} & 0 & 0 \\
         a_{21} & -\frac{1}{2} & 0 \\
         -\frac{1}{2} & \frac{3}{2} & -\frac{1}{2} \\
        \end{array}
        \right].
\end{equation}
\begin{figure}[h!]
    \centering\includegraphics[width=0.5\textwidth]{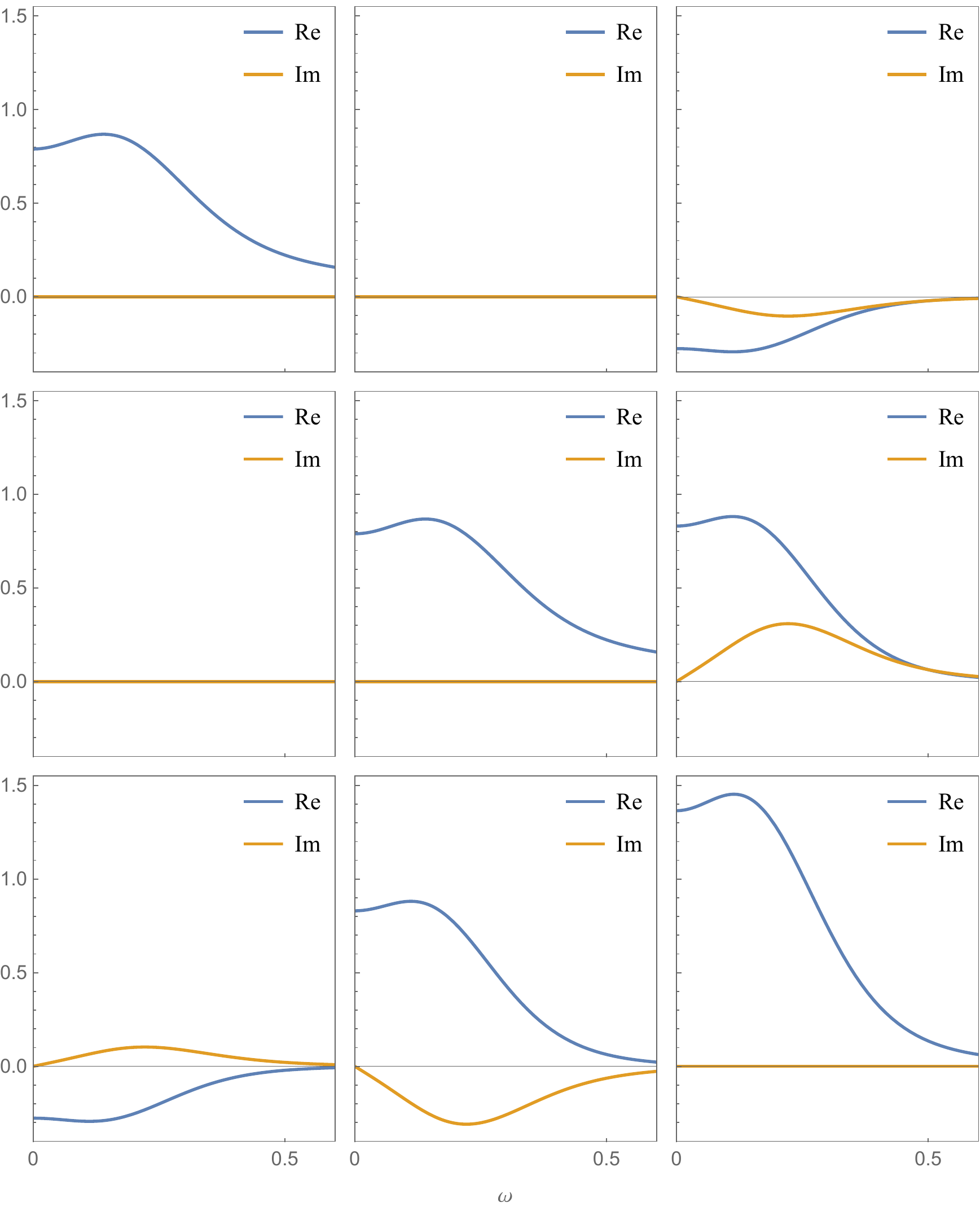}
    \caption{\label{fig:CSD_y_0_normalised}
        Simulated BOLD cross-spectral density of the system in \cref{fig:3_variables_network_graph}, generated via the forward model in \cref{eq:CSD_BOLD_forward}.
        The effective connectivity matrix is presented in \cref{eq:A_3D_example}.
        This figure only shows the special case where $a_{21}=0$ but using different values of $a_{21}$ would generate different sets of nine plots.
        Each plot corresponds to a pair of state variables and, if we chose a specific value for the frequency $\omega$, we would obtain a $3 \times 3$ matrix (the cross-spectral density matrix at that specific frequency).
        The three plots on the diagonal represent the power spectral densities, which take real values (zero imaginary part).
        The same canonical hemodynamic response function defined in \cref{eq:HRF_canonical_Fourier} is used for all variables in this example, although spectral DCM employs the BOLD balloon model with region-specific biophysical parameters.
        The power-law parameter of the endogenous fluctuations and of the observation noise are set as in \cref{eq:power_law_decay_2D_example}.
        For simplicity, we have normalised the rows and columns of the cross-spectral density so that integrating over all frequencies generates the correlation matrix instead of the covariance matrix.
    }
\end{figure}

As before, the strength of the directed connection from $x_1$ to $x_2$ is set via the parameter $a_{21}$.
The third neuronal state variable receives an inhibitory influence from $x_1$ and an excitatory influence from $x_2$.
All three state variables have the same negative self-connections represented by the diagonal elements of $A$.
The resulting cross-spectral density is plotted in \cref{fig:CSD_y_0_normalised}, assuming the same canonical HRF defined in \cref{eq:HRF_canonical_Fourier} and the same power-law parameters as in \cref{eq:power_law_decay_2D_example}.
The figure only shows the special case where $a_{21}=0$ but different values of $a_{21}$ would generate different sets of nine plots.

\begin{figure}
    \centering\includegraphics[width=0.5\textwidth]{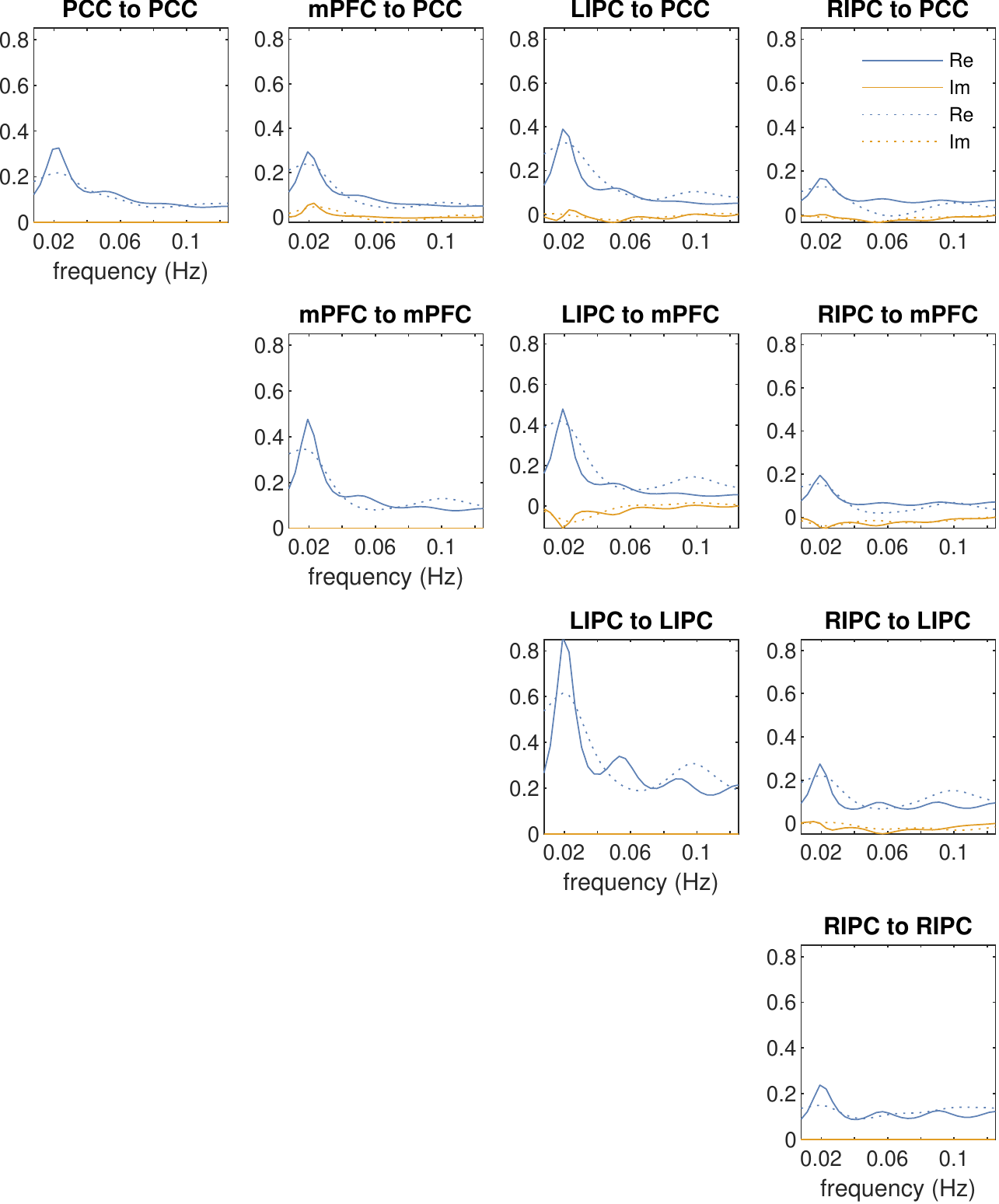}
    \caption{\label{fig:CSD_real_data}
        BOLD cross-spectral density obtained via spectral DCM analysis of a real resting-state fMRI dataset.
        The dashed lines represent the predicted cross-spectral density and the solid lines the observed ones.
        Reproduced with permission from Chapter 38 of the SPM12 manual \cite{johnashburner2020SPM12Manual}.
    }
\end{figure}

It is instructive to qualitatively compare the simulated cross-spectral density plots in \cref{fig:CSD_y_0_normalised} with the empirical plots in \cref{fig:CSD_real_data}, obtained by fitting the spectral DCM model to a real resting-state dataset \cite{razi2015ConstructValidationDCM}.
Note that the empirical plots correspond to a system with four neuronal variables instead of three.
Here, a four regions default mode network (DMN) is modelled with posterior cingulate cortex (PCC), medial prefrontal cortex (mPFC) and bilateral inferior parietal cortices (IPC) as its nodes.
We used a fully connected model where each state variable (or node) is connected to every other state variable. 
The figure is reproduced from Chapter 38 of the SPM12 manual \cite{johnashburner2020SPM12Manual}, which provides the link to the data and a step-by-step tutorial to replicate the DCM specification and estimation results using the SPM graphical interface.
Despite the clear differences due to different data and parameter setting, the empirical cross-spectral density plots also feature a single large peak at low frequencies, followed by a decay at larger frequencies (with smaller fluctuations).
If the data is too noisy, due to head-motion and various physiological artifacts, additional large peaks may appear at higher frequencies.
These empirical plots can be visualised using the review function (\texttt{spm\_dcm\_fmri\_csd\_results()}) in SPM.
Another DCM diagnostics function (\texttt{spm\_dcm\_fmri\_check()}) in SPM also reports the percentage of variance explained ($R^2$), which is a useful performance metric to judge the quality and success of the model fit to the data (see \cref{fig:spm_dcm_fmri_check}).

\section{Effective and functional connectivity}
\label{sec:EC_vs_FC}
\begin{figure}[h]
    \centering\includegraphics[width=0.5\textwidth]{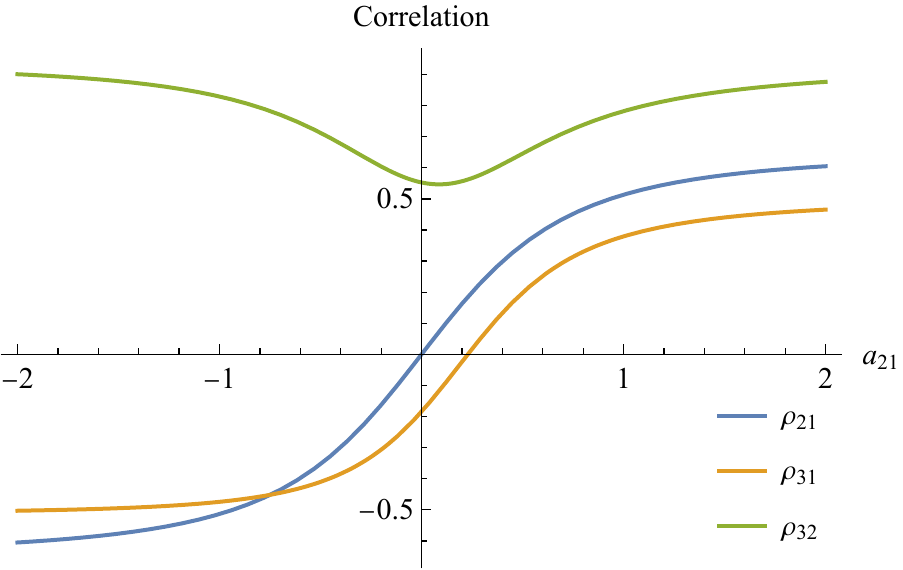}
    \caption{\label{fig:corr_y}
        Simulated BOLD functional connectivity of the system in \cref{fig:3_variables_network_graph}, as a function of the effective connectivity parameter $a_{21}$.
        The three curves corresponding to the correlation values are obtained analytically by integrating the cross-spectral density shown in \cref{fig:CSD_y_0_normalised} over all frequencies.
        Changing a single effective connectivity parameter, $a_{21}$, has a global impact across all pairwise functional connectivity values in the network.
        Both correlations $\rho_{21}$ and $\rho_{31}$ increase monotonically with $a_{21}$, but this is not the case for $\rho_{32}$.
        Therefore, there is no one-to-one mapping between effective and functional connectivity that holds in general.
    }
\end{figure}

\textls[15]{
The main difference between effective and functional connectivity is that the former characterises the interaction between neuronal state variables, while the latter describes statistical dependence between the observed variables (\eg the BOLD signals in fMRI).
Given its logical and biological precedence, effective connectivity can be used to derive functional connectivity.
Specifically, the functional connectivity matrix of the observed time series can be obtained by integrating the cross-spectral density over all frequencies.
The reason will become clear in the following sections and a mathematical proof will be given in \cref{eq:FC_from_CSD}.
Let's first return to the simulated example depicted in \cref{fig:3_variables_network_graph} to understand the relevant implications.
The effective connectivity matrix of the system is given in \cref{eq:A_3D_example}.
Integrating the cross-spectral density in \cref{fig:CSD_y_0_normalised} over all frequencies gives us the $3 \times 3$ symmetric correlation matrix $R$, typically used to quantify the functional connectivity\footnote{Usually, integrating the cross-spectral density over all frequencies would produce the covariance matrix; normalising its rows and columns would then generate the correlation matrix as a second step. In \cref{fig:CSD_y_0_normalised}, we have already normalised the rows and columns of the cross-spectral density so that integrating over all frequencies generates the correlation matrix directly. The nine integrals produce real numbers because all the imaginary parts are `odd functions' so their integrals are equal to zero.}:
\begin{equation}
  R =
  \left[ {\begin{array}{ccc}
    1           & \rho_{21} & \rho_{31}\\
    \rho_{21}   & 1         &  \rho_{32}\\
    \rho_{31}   & \rho_{32} & 1
  \end{array} } \right].
\end{equation}
Crucially, all three pairwise correlations ($\rho_{21}, \rho_{31}, \rho_{32}$) explicitly depend on the effective connectivity parameter $a_{21}$, as illustrated in \cref{fig:corr_y} (for the analytic solutions, see \cref{sec:correlation_matrix_3by3_analytic}).
}

This shows that a local variation in a single effective connectivity parameter in the $A$ matrix can have a global impact across all functional connectivity values in the network.
Let's examine each pair to unpack some of the many nuances involved.
First, the symmetric nature of the correlation matrix requires that $\rho_{kj}=\rho_{jk}$, even if $a_{kj} \neq a_{jk}$.
This is a key difference between functional and effective connectivity: only the latter is directed and is able to differentiate between two bilateral connections.

Both correlations $\rho_{21}$ and $\rho_{31}$ increase monotonically with $a_{21}$, but this is not the case for $\rho_{32}$.
Therefore, there is no one-to-one mapping between effective and functional connectivity that holds in general \cite{park2013StructuralFunctionalBrain}.
This poses a challenge for spectral DCM because it makes model fitting an ill-posed problem with multiple potential solutions.
Technically, this problem is mitigated by using the cross spectral density that implicitly contains information about functional connectivity over all lags (we will unpack this below) and by using priors on the solutions implicit in the functional form of the DCM.
Although model fitting remains an ill-posed problem, these two additional constraints allow spectral DCM to find better solutions, such that the model can reproduce a larger set of statistical relationships between the observed time series.
One way to appreciate the amount of additional information and constraints provided by the cross-spectral density over the zero-lag correlation is to inspect \cref{fig:CSD_y_0_normalised} again.
Zero-lag correlation only measures the area under the curves in the nine plots, regardless of their detailed shapes.
On the other hand, spectral DCM fits all values taken by the cross-spectral density curves at different frequencies.
This requirement narrows the parameter space is a useful way (note that the area under the curve is still reproduced as a consequence).

The sign of $\rho_{31}$ (representing a positive or a negative correlation) either matches or contradicts the sign of the underlying effective connectivity $a_{31}$ (representing an excitatory or an inhibitory connectivity), depending on how we set $a_{21}$.
There is a value of $a_{21}$ such that $\rho_{31}=0$, even though the underlying effective connectivity $a_{31}$ is negative (\ie inhibitory).
Again, this shows that each pairwise functional connectivity value is a summary statistic (global property) of the system: given a pair of variables $j$ and $k$, the correlation value $\rho_{kj}$ does not only depend on the corresponding effective connectivity parameter $a_{kj}$ but, potentially, on all the entries of the $A$ matrix.

While here we have only discussed its dependence on the effective connectivity, functional connectivity also depends on the parameters characterising the endogenous fluctuations, the observation function, and the observation noise.
This is because all these parameters appear in the forward model for the cross-spectral density derived in \cref{eq:CSD_BOLD_forward} and, in turn, determine the correlation matrix of the system. 
In fMRI, specific and reproducible spatial patterns of functional connectivity could simply arise from specific and reproducible variations in the hemodynamic response across brain regions---even in the absence of interregional effective connectivity \cite{rangaprakash2018HemodynamicResponseFunctiona}.
When comparing groups, differences in BOLD functional connectivity may reflect differences in the vasculature rather than in effective connectivity (or both), \eg due to ageing or to a neurodegenerative or psychiatric disease \cite{tsvetanov2020SeparatingVascularNeuronal}. 
%More realistically, functional connectivity would reflect both the underlying effective connectivity and regional variations in hemodynamics and observation noise.
%For example, subcortical nuclei and cortical areas near the temporal pole have lower signal-to-noise ratios compared to other cortical regions.
%These differences in observation noise likely confound the functional connectivity of these areas.
This important point is further discussed in \cite{friston2011FunctionalEffectiveConnectivity}, where it is shown that the correlation values are also influenced by different levels of observation noise: ``one can see a change in correlation by simply changing the signal-to-noise ratio of the data.
This can be particularly important when comparing correlations between different groups of subjects. For example, obsessive compulsive patients may have a heart rate variability that differs from normal subjects. This may change the noise in observed hemodynamic responses, even in the absence of neuronal differences or changes in effective connectivity.''
In fact, separating these confounding factors from the effective connectivity was one of the main motivations for the development of DCM.
This is not to say that DCM is without issues: fitting a large model with many parameters---using the limited amount of data available in typical fMRI studies---is not guaranteed to produce optimal estimates.
We'll discuss this and other limitations in \cref{sec:assumptions}.

Functional connectivity faces yet another challenge. 
Even though we observed that any variations in effective connectivity can propagate through the network and affect the correlation among several brain areas, one could hope that the pair of variables showing the largest functional connectivity change would coincide with the pair evincing the largest effective connectivity change.
Alas, this is also not guaranteed.
When $a_{21}$ increases from zero to one, $R$ shows the following changes:
\begin{equation}
    \Delta R = \left[
        \begin{array}{ccc}
         0 & 0.513367 & 0.379316 \\
         0.513367 & 0 & 0.782089 \\
         0.379316 & 0.782089 & 0 \\
        \end{array}
        \right].
\end{equation}
The largest functional connectivity change is observed in $\rho_{32}$, not in $\rho_{21}$ as we might have expected.
The fact that the variable pairs showing the largest changes in functional connectivity do not necessarily coincide with the pairs with the largest changes in effective connectivity may hinder the use of functional connectivity as a quick and easy way for selecting regions of interest in DCM studies.

On the other hand, functional connectivity has proven valuable for fingerprinting, that is, identifying an individual based on their brain activity \cite{finn2015FunctionalConnectomeFingerprinting}.
It is also useful in studies aiming to differentiate between two groups or conditions (by detecting statistically significant changes in correlation between observed time series), rather than in identifying which effective connections between brain regions underlie that change.
Moreover, being a global property of the system, each pairwise functional connectivity value naturally captures higher-order interactions, which is a topic of growing interest in complex systems, network science, and neuroimaging \cite{benson2016HigherorderOrganizationComplex,rosas2022DisentanglingHighorderMechanisms}.

So far, we have only examined the zero-lag correlation matrix because it is widely used to quantify the functional connectivity in neuroimaging.
However, spectral DCM doesn't just explain the zero-lag correlation but also the cross-correlations between all variables at all time lags.
To understand this point, in the next section, we will invoke the elegant Wiener-Khinchin theorem, which links the cross-covariance function to the cross-spectral density.

\subsection{Autocovariance and cross-covariance functions}
\label{sec:autocovariance}
Let's briefly revisit the role of the self-connections in the presence of the endogenous fluctuations.
As discussed in \cref{sec:effective_connectivity}, a self-connection determines the rate of decay of a variable.
It can also be understood as quantifying the memory of a variable, \ie whether inputs have a short-lived or long-lasting impact on its activity.
Large negative self-connections reflect short memory: the variable ``forgets'' its past quickly and responds promptly to any new inputs.
On the contrary, small negative values reflect long memory: the neuronal variable integrates and smooths out the endogenous fluctuations and other inputs, resulting in slower oscillations and lower frequencies.

The concept of a system having memory can be quantified via the \emph{autocovariance function}.
For a deterministic signal $z(t)$, there is a simple intuition: the autocovariance function at a time lag $\Delta t$ measures the similarity (\ie sample covariance) between the time series $z(t)$ and a shifted version of itself, that is, $z(t+\Delta t)$.
The agreement is perfect when there is no shift ($\Delta t=0$) and it typically decreases with longer time lags, unless the signal is constant or periodic.
However, spectral DCM is concerned with stochastic (non-deterministic) processes, as we discussed in \cref{sec:PSD_and_CSD} and illustrated in \cref{app:stochastic_processes}.
Let's consider the single stochastic neuronal variable $x_1$.
At any given time point, $x_1(t)$ is not a number but a random variable.
If, after a time interval $\Delta t$, the random variable $x_1(t+\Delta t)$ is still positively correlated with its previous state $x_1(t)$, the autocovariance between the two time points would be positive.
If, as the time interval further increases, $x_1$ forgets its past state and become independent of it, the autocovariance would become zero at that point (for BOLD signals, time intervals or time lags are always multiples of the repetition time).
%\footnote{This is the ensemble covariance rather than the sample covariance used in the deterministic example above.} 
In summary, the autocovariance function measures the covariance between the states of the same stochastic process at two different points in time.
For the stationary processes considered here, the autocovariance only depends on the time interval $\Delta t$:
\begin{equation} \label{eq:autocov_general}
    \sigma_{11}(\Delta t) = \textup{cov}[x_1(t),x_1(t+\Delta t)].
\end{equation}
Dividing the autocovariance function by $\sigma_{11}(0)$ would produce the commonly used \emph{autocorrelation function}, whose values are normalised to the $[-1,1]$ interval.
%Note that the autocovariance of a stationary process is always an \emph{even} function, \ie it takes the same values for $\Delta t$ and $-\Delta t$.

We can compute the autocovariance explicitly in the simple scenario of \cref{eq:self-connection_stochastic}, where $x_1$ is driven by endogenous fluctuations only, \ie it doesn't receive any other inputs.
For simplicity, assume that the endogenous fluctuations were modelled as a white-noise process with $\alpha_{v_1}=1$ and $\beta_{v_1}=0$.
The equation would then describe an Ornstein-Uhlenbeck process that has the following autocovariance function \cite{vatiwutipong2019AlternativeWayDerive}:
\begin{equation} \label{eq:autocov_Ornstein}
    \sigma_{11}(\Delta t) = \frac{e^{a_{11} |\Delta t|}}{-2a_{11}}.
\end{equation}
This confirms the intuition above: large negative values of $a_{11}$ correspond to short memory, in this case, the decay is exponential.

\begin{figure}
    \centering\includegraphics[width=0.5\textwidth]{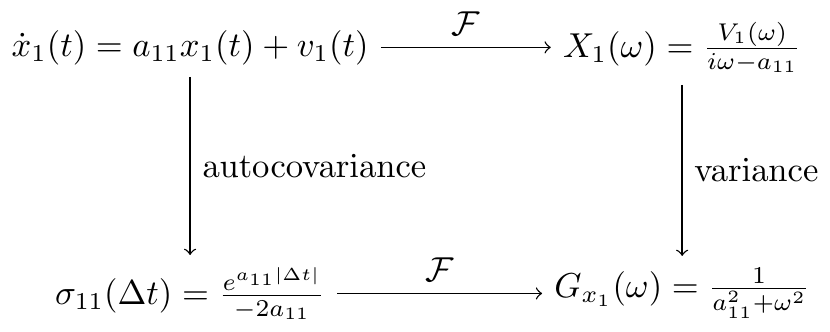}
    \caption{ \label{fig:arrow_diagram}
        The two equivalent representations of the power spectral density.
        Starting from the Ornstein-Uhlenbeck process as an example (top left), the power spectral density can either be computed as the variance of the Fourier transform (top right, then bottom right), or as the Fourier transform of the autocovariance function (bottom left, then bottom right).
        The former uses the definition of the power-spectral density; the latter uses the Wiener-Khinchin theorem.
        In the Ornstein-Uhlenbeck process, the endogenous fluctuations ($v_1(t)$) are modelled as a white-noise process.
        Its Fourier transform, denoted as $V_1(\omega)$, is a Normal random variable for each frequency $\omega$.
    }
\end{figure}

There is another reason why the autocovariance function is relevant for spectral DCM and, more generally, for DCM applications to electroencephalography (EEG) and magnetoencephalography (MEG).
These models and modalities are concerned with the power spectral density of signals, and the Wiener-Khinchin theorem proves that the power spectral density of a stationary process is the Fourier transform of its autocovariance function.
In the case of $x_1$, we get
\begin{equation} \label{eq:power_spectrum_x1_from_autocovariance}
    G_{x_1}(\omega) = \mathcal{F}\{ \sigma_{11}(\Delta t) \} = \mathcal{F}\{ \frac{e^{a_{11} |\Delta t|}}{-2a_{11}} \} =\frac{1}{a_{11}^2 + \omega^2}.
\end{equation}
This offers an alternative way to compute the power spectral density compared to the definition given in \cref{sec:PSD_and_CSD}.
The two equivalent representations are often portrayed with arrow diagrams in spectral DCM papers \cite{friston2014DCMRestingState,razi2016ConnectedBrainCausality}.
Here, a similar diagram is provided in \cref{fig:arrow_diagram} using the Ornstein-Uhlenbeck process as an example.

In the presence of multiple state variables, the autocovariance function is replaced by the more general \emph{cross-covariance function}
\begin{equation} \label{eq:cov_func_x}
    \Sigma_{x}(\Delta t) = \textup{cov}(\bm{x}(t),\bm{x}(t+\Delta t)).
\end{equation}
For any time lag $\Delta t$, the cross-covariance function produces a $N \times N$ matrix with entries
\begin{equation}
    [\Sigma_{x}(\Delta t)]_{jk} = \sigma_{jk}(\Delta t) = \textup{cov}(x_j(t),x_k(t+\Delta t)).
\end{equation}
Each diagonal element in this matrix is the autocovariance function of a single neuronal variable, introduced in \cref{eq:autocov_general}.
The off-diagonal elements are the cross-covariance functions between different state variables.
For deterministic processes, the visual intuition is analogous to the autocovariance case: for each pair of variables $x_j$ and $x_k$, the cross-covariance function $\sigma_{jk}(\Delta t)$ measures the similarity between the time series $x_j(t)$ and a shifted version of $x_k$ by a time lag $\Delta t$.
The analogy with memory can then be used to extend the intuition to stochastic processes. 
%Unlike the autocovariance, the cross-covariance function is not necessarily an even function so it can take different values for $\Delta t$ and $-\Delta t$.
%Normalising the cross-covariance function by $...$ generates the more commonly known \emph{correlation function} and \emph{correlation matrix}, whose values are normalised to the $[-1,1]$ interval.
When there is no time lag ($\Delta t=0$), the cross-covariance function produces the covariance matrix $\Sigma_{x}(0)$, which is a symmetric matrix because $\sigma_{jk}(0)=\sigma_{kj}(0)$.
After normalisation, this yields the correlation matrix that is used to quantify the functional connectivity in neuroimaging. 
%However, when there is a time lag ($\Delta t \neq 0$), the lagged covariance matrix $\Sigma_{x}(\Delta t)$ is anti-symmetric since $\sigma_{jk}(\Delta t)=-\sigma_{kj}(\Delta t)$ for stationary processes.

\subsection{Cross spectral density and functional connectivity}
\label{sec:CSD_Khinchin}
We can again invoke the Wiener-Khinchin theorem and obtain the cross-spectral density as the Fourier transform of the cross-covariance function:
\begin{equation} \label{eq:CSD_x_from_covariance}
    G_x(\omega) = \mathcal{F}\{ \Sigma_x(\Delta t) \}.
\end{equation}
We previously noted that the off-diagonal elements of the cross-spectral density matrix are complex numbers and that their definition provided in \cref{eq:CSD_neuronal} doesn't lend itself to an intuitive understanding.
However---now armed with the Wiener-Khinchin theorem---we gain another perspective.
Each off-diagonal element is equivalent to the Fourier transform of the cross-covariance function between two variables.
By modelling the cross-spectral density, spectral DCM doesn't just capture the zero-lag correlations but also the cross-correlations between all variables at all time lags.
This is true for both the hidden neuronal variables and the observed (\eg BOLD) variables.
The equivalent of \cref{eq:CSD_x_from_covariance} for the observed variables is:
\begin{equation} \label{eq:CSD_y_from_covariance}
    G_y(\omega) = \mathcal{F}\{ \Sigma_y(\Delta t) \}.
\end{equation}

We opened \cref{sec:EC_vs_FC} by stating that the functional connectivity matrix can be obtained by integrating the cross-spectral density over all frequencies.
To understand why, we need to invert \cref{eq:CSD_y_from_covariance} using the (inverse) Fourier transform
\begin{equation}
     \Sigma_y(t) = \mathcal{F}^{-1}\{ G_y(\omega) \} = \int^{\infty }_{-\infty } G_y(\omega) e^{i \omega t} \diff \omega,
\end{equation}
which retrieves the cross-covariance function from the cross-spectral density $G_x(\omega)$.
When $t=0$, we have the desired result
\begin{equation} \label{eq:FC_from_CSD}
     \Sigma_y(0) = \mathcal{F}^{-1}\{ G_y(\omega) \} = \int^{\infty }_{-\infty } G_y(\omega) \diff \omega.
\end{equation}
The correlation matrix $R$ is finally obtained by normalising the covariance matrix.

\section{Assumptions and limitations}
\label{sec:assumptions}
\begin{description}
    \item[State-space modelling]{
    The foundational hypothesis of the DCM framework is that the system of interest can be modelled using a state-space approach.
    This formulation separates the equations describing the temporal evolution of unobserved variables (\eg the neuronal activity) from those describing the observed variables (\eg the BOLD).
    Such a separation is important because some of the model parameters have a direct interpretation in terms of effective connectivity between unobserved neuronal populations rather than statistical dependency between observed variables.
    A subtle and important consequence of the distinction between functional effective connectivity is that one can only estimate recurrent or self inhibition using a state space model.
    This is because the correlation of a time series with itself is always one (and the variance is always positive).
    Assessing self connectivity in terms of excitability or disinhibition of a node can be empirically important (\eg in estimating changes in excitation-inhibition balance or condition-specific changes in the `gain' of a particular source or region).
    %State space modelling is also necessary to estimate directed connectivity (\eg distinct changes in forward and backward connections between two nodes in a cortical hierarchy). 
    }
    \item[Continuous-time formulation]{
    Like most DCM approaches, spectral DCM treats time as a continuous quantity.
    Representing time as a sequence of discrete steps may seem more natural, especially for fMRI, where the observations are recorded at evenly spaced time intervals with a relatively low sampling rate.
    However, DCM also models the neuronal activity, which unfolds at a much faster time scale and in an asynchronous manner.
    These two features can be naturally modelled using differential equations in continuous time.
    The discrete formulation usually converges to the continuous one with faster acquisition times, as is the case for EEG and MEG.
    }
    \item[Separation of time scales and macroscopic modelling]{
    Spectral DCM assumes that a single macroscopic neuronal variable can capture the essential dynamical properties of a population of neurons.
    This is not just a pragmatic way to avoid billions of equations representing individual neurons.
    It is an approach based on the separation of time scales that has a long history in dynamical systems theory \cite{carr1981ApplicationsCentreManifold,haken1983NonlinearEquationsSlaving}.
    The assumption is that the neuronal activity can be separated into slow and fast modes: the fast modes decay quickly so that the long-term behaviour of the system can be described by a few slow modes, or even a single one.
    Mathematically, the assumption (often satisfied in real systems) is that only a small number of eigenvalues of the Jacobian are near zero, while the rest are large and negative \cite{friston2011NetworkDiscoveryDCM,friston2021ParcelsParticlesMarkov}.
    Macroscopic modelling also relies on the mean-field assumption that the dynamics of one region are determined by the mean activity in another \cite{deco2008DynamicBrainSpiking}.
    %The intuition is that, if each neuron in one region receives inputs from a sufficiently large number of neurons in another, the collective influence is assumed to be equal to the mean over all neurons.
    Since the mean activity is dominated by the slow modes, it is possible to build a compact macroscopic model where only the slow modes are communicated among brain regions, whereas the fast endogenous fluctuations only affect the local activity.
    This is precisely the neuronal model in \cref{eq:neuronal_model_full}.
    It is important to note that, despite being grounded in dynamical and complex systems theory, this model is an abstraction.
    Biological details are necessarily omitted to enable the analytic treatment and faster numerical computations.
    A detailed critical review of the biophysical and statistical foundations of DCM is provided in \cite{daunizeau2011DynamicCausalModelling}.
    }
    \item[Stationarity]{
    Among the DCM variants, spectral DCM has the most direct conceptual and analytical links to functional connectivity, which we have examined in \cref{sec:EC_vs_FC}.
    Both methods can be read as assuming that the observed processes are weakly stationary, \ie that their covariance remains unchanged over the length of the experiment \cite{liegeois2017InterpretingTemporalFluctuations}.
    In the case of functional connectivity, this assumption arises because the covariance is used as a proxy for connectivity, after being normalised to obtain the correlation matrix.
    Therefore, assuming weak stationarity is equivalent to assuming that the functional connectivity remains unchanged.
    Similarly, in spectral DCM, the stationarity assumption allows one to interpret the effective connectivity as remaining unchanged over the length of the time series \cite{moran2009DynamicCausalModels,friston2014DCMRestingState}.
    This is because the effective connectivity is inferred from the observed cross-spectral density, \ie the Fourier transform of the cross-covariance function (see \cref{sec:CSD_Khinchin}), which remains unchanged under stationarity assumptions.
    However, in contrast to functional connectivity, spectral DCM only treats the cross-covariance of the observed time series as a means to an end, where the end is to infer the effective connectivity between neuronal variables (and various other model parameters).
    In other words, spectral DCM looks ``under the hood'' of functional connectivity and beyond the observed variables.
    
    The stationarity assumption can certainly be challenged and it is generally untenable in task experiments; nonetheless, it is widely adopted in resting-state studies.
    Stationarity is not just typically assumed in functional connectivity and spectral DCM, but also in Granger causality analysis \cite{granger1969InvestigatingCausalRelations,seth2013GrangerCausalityAnalysis}, transfer entropy \cite{bossomaier2016IntroductionTransferEntropy} and autoregressive modelling \cite{liegeois2017InterpretingTemporalFluctuations}.
    That said, Granger causality and information-theoretic methods have been adapted for non-stationary processes \cite{dhamala2008EstimatingGrangerCausality,lizier2008LocalInformationTransfer,gomez-herrero2015AssessingCouplingDynamics,wollstadt2014EfficientTransferEntropy,palus2017LinkedDynamicsWavelet} and time-varying approaches to functional connectivity analysis have been rapidly gaining popularity \cite{lurie2020QuestionsControversiesStudy,novelli2022MathematicalPerspectiveEdgecentric}.
    A time-varying extension of spectral DCM has also been developed \cite{park2018DynamicEffectiveConnectivity}.
    All of these methods relax the stationarity assumption and allow the statistical properties of the system to change over the course of the experiment.
    Practically, in DCM, one appeals to something called an adiabatic approximation: that effective connectivity is constant over a small timescale but can change at longer timescales.
    This means that one can apply spectral DCM to short segments of data and then examine (or model) slow fluctuations in effective connectivity \cite{jafarian2021AdiabaticDynamicCausal,rosch2019FunctionalGenomicsEpilepsy,zarghami2020DynamicEffectiveConnectivity}.
    }
    \item[Linearity]{
    The second assumption that spectral DCM (partially) shares with functional connectivity is linearity, although nonlinear extensions of both methods exist \cite{stephan2008NonlinearDynamicCausal,kraskov2004EstimatingMutualInformation}.
    In DCM for fMRI, the use of a linear random differential equation to model the neuronal activity is motivated by the separation of time scales, whereby the neuronal variables represent the slow modes of the system, which are assumed to be linearly coupled, while the endogenous random fluctuations represent the fast modes \cite{friston2011NetworkDiscoveryDCM}.
    Spectral DCM also extends the linearity assumption to the observation function, which can be a nonlinear function of time and frequency (as in \cref{eq:HRF_canonical_Fourier}) but is linearly convolved with the neuronal activity (\cref{eq:HRF_convolution_single}).
    %Here, we've often referred to the BOLD signal as the observed variable because spectral DCM has historically been applied to fMRI.
    %However, the mathematical treatment is general and replacing the HRF with a different observation function allows its application to different modalities.
    In \cref{sec:endogenous_fluctuations}, we have observed that deterministic linear models have a limited scope since they can only describe a system that converges to equilibrium or generates a sequence of identical oscillations.
    Adding stochastic terms to the linear differential equations allows for a richer repertoire of behaviours.\footnote{Interestingly, a linear model driven by intermittent inputs can even replicate the switching behaviour of the chaotic Lorentz system \cite{brunton2017ChaosIntermittentlyForced}.}
    }
    \item[Stochastic fluctuations]{
    The addition of a stochastic term to a dynamical system is traditionally used to model noise, often assumed to be white and temporally uncorrelated.
    The underlying assumption is that the noise is due to physical processes operating at a much faster time scale than the state variables, \eg microscopic thermal fluctuations.
    Spectral DCM relaxes this assumption and allows the endogenous fluctuations to be temporally correlated, with a spectrum following a power-law decay in the frequency domain (see \cref{sec:endogenous_fluctuations} to see how this form includes white noise as a special case).
    This is easily motivated by noting the endogenous fluctuations are themselves generated by dynamical processes within the source or region of interest.
    The ensuing temporal autocorrelation makes the endogenous fluctuations differentiable and smooth, in line with their characterisation as mixtures of fast modes of a dynamical system \cite{friston2011NetworkDiscoveryDCM}.
    %Indeed, the power-law decay is typical of linear dynamical systems.
    %For example, an Ornstein-Uhlenbeck process has the Lorentzian power spectrum as in \cref{eq:power_spectrum_x1_from_autocovariance}.
    %The endogenous fluctuations are temporally autocorrelated but spatially independent
    }
    \item[Gaussianity]{
    In line with most DCM approaches, spectral DCM assumes a Gaussian distribution for the prior over model parameters (the non-negative parameters are transformed using the natural logarithm and are assumed to follow a log-Normal distribution).
    %The neuronal states are not inferred directly but, if we assume that the stochastic process that represents them has a stationary Gaussian distribution, then the second-order statistic fully describe it.
    This enables a fast Bayesian inversion scheme called variational Laplace \cite{zeidman2022PrimerVariationalLaplace,friston2007VariationalFreeEnergy}.
    As for most of the assumptions listed in this section, the Gaussian hypothesis can also be relaxed and other (albeit more computationally intensive) inversion schemes can be used instead, \eg Markov chain Monte Carlo methods \cite{friston2007VariationalFreeEnergy,aponte2022IntroductionThermodynamicIntegration,xie2023SpectralSamplingAlgorithm}.
    }
    \item[Many-to-one mapping]{
    There is no one-to-one mapping between effective and functional connectivity (or cross-spectral density) that holds in general.
    This is a challenge for spectral DCM because it makes model inversion an ill-posed problem with multiple potential solutions, in the absence of any constraints on the way data are generated.
    As with all ill-posed problems, this is addressed by placing prior constraints on the explanations in the form of a model and prior densities over model parameters.
    When one does not know which priors to use, a weighted average of plausible priors is often performed in DCM analysis using Bayesian model averaging \cite{hoeting1999BayesianModelAveraging}), where each set of priors corresponds to a separate model.
    A related challenge is that one cannot always use functional connectivity to identify the regions of interest to study using DCM.
    As we saw in \cref{sec:EC_vs_FC}, not even the regions showing the largest differences in correlation are guaranteed to coincide with the regions with a change in effective connectivity.
    }
    \item[Computational complexity]{
    Despite the simplifying assumptions mentioned above, the computational complexity of spectral DCM limits the number of brain regions that can be studied in reasonable time.
    Selecting the regions of interest requires more upfront work than in functional connectivity analysis, which can be quickly performed across the whole brain.
    Given the large size of the parameter space, the specification of the model has been noted as a conceptual issue for DCM in the past \cite{friston2013AnalysingConnectivityGranger}.
    That said, a more recent theoretical advance now enables the exploration of a large model space using Bayesian model reduction \cite{seghier2013NetworkDiscoveryLarge}.
    Alternatively, one can introduce further assumptions to winnow the parameter space.
    Using functional connectivity to place prior constraints on the eigenvectors of the effective connectivity matrix enables spectral DCM analyses with dozens of brain regions \cite{razi2017LargescaleDCMsRestingstate}.
    In fMRI, ignoring the spatial variability of the hemodynamics and removing the separation between hidden and observed variables leads to the \emph{regression DCM} scheme, which can analyse hundreds of regions in minutes \cite{frassle2021RegressionDynamicCausal}.
    However, this method forgoes the state-space formulation and can be understood as a Bayesian multivariate regression in the frequency domain.
    }
\end{description}

%\section*{Code availability}
%

\section*{Author Contributions}
\noindent Leonardo Novelli: Conceptualization; Formal analysis; Investigation; Software; Visualization; Writing --- original draft.

\noindent Karl Friston: Conceptualization; Writing --- review \& editing.

\noindent Adeel Razi: Conceptualization; Funding acquisition; Supervision; Writing --- review \& editing.

\section*{Acknowledgements}
L.N. is funded by the Australian Research Council (Ref: DP200100757).
A.R. is funded by the Australian Research Council (Refs:  DE170100128  and  DP200100757) and Australian National Health and Medical Research Council Investigator Grant (Ref: 1194910).
A.R. is affiliated with The Wellcome Centre for Human Neuroimaging supported by core funding from Wellcome [203147/Z/16/Z].
A.R. is a CIFAR Azrieli Global Scholar in the Brain, Mind \& Consciousness Program.
K.F. is funded by the Wellcome Centre for Human Neuroimaging (Ref: 205103/Z/16/Z) and a Canada-UK Artificial Intelligence Initiative (Ref: ES/T01279X/1).

\nolinenumbers

%% If you have bibdatabase file and want bibtex to generate the
%% bibitems, please use

%\bibliographystyle{elsarticle-harv} 
%\bibliography{bibliography}

%% else use the following coding to input the bibitems directly in the TeX file.

\newpage

%% The Appendices part is started with the command \appendix;
%% appendix sections are then done as normal sections
\appendix

\section{Stochastic processes}
\label{app:stochastic_processes}
\begin{figure*}[p]
    \centering\includegraphics[width=\textwidth]{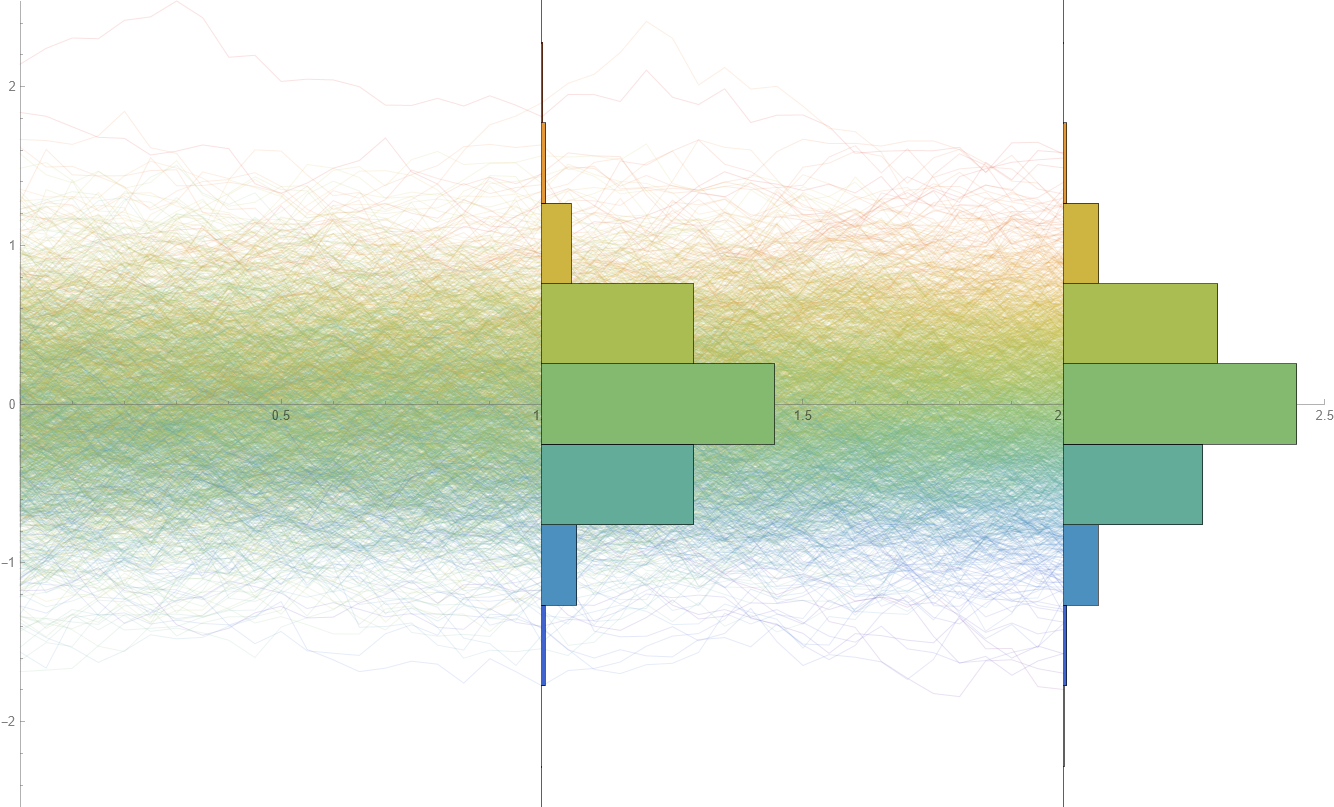}
    \caption{\label{fig:OU_process_slice_distributions}
        Illustration of an ensemble of realisations of an Ornstein-Uhlenbeck process.
        Each curve is a different realisation, independent of the others.
        The process is sliced at two time points $t=1$ and $t=2$.
        Although the individual realisations (curves) take different values at these two time points, they collectively preserve the same Gaussian distribution.
        This property is known as stationarity and can be assumed when using  spectral DCM.
        Non-stationary processes have different distributions at different time points and would require different models that allow for time varying parameters, \eg stochastic DCM or adiabatic DCM.
    }
\end{figure*}

A stochastic process is a sequence of random variables.
If $x$ is a stochastic process indexed by time, then $x(t)$ is not a single number but a random variable with a given probability distribution.
Intuitively, the process is stationary if, when we collect many realisations (known as a statistical \emph{ensemble}) and plot a histogram of their values at different time points, we will obtain the same distribution.
\cref{fig:OU_process_slice_distributions} illustrates this concept using a stationary Ornstein-Uhlenbeck process and showing two histograms of the ensemble at two different time points.
Although the individual realisations (curves) take different values, they collectively preserve the same Gaussian distribution.

\section{Real and imaginary parts of the cross-spectral density}
\label{app:real_imaginary_CSD}
The Fourier transform produces complex numbers, which can be described in Cartesian coordinates (real and imaginary parts on the complex plane) or in polar coordinates (amplitude and argument).
Polar coordinates offer an intuitive interpretation: the amplitude of the Fourier transform denotes how strongly a given frequency is represented in the signal, while the argument indicates how much the contribution of this frequency is phase-shifted.
In Cartesian coordinates, the interpretation is less intuitive: the real and imaginary parts of the cross-spectral density are the Fourier transforms of the even and odd parts of the cross-covariance function.
What are the even and odd parts?
Before giving the mathematical definition, consider the sine and cosine functions.
The cosine function produces the same output regardless of the sign of its input, that is, $\cos(t)=\cos(-t)$.
Functions with this property are symmetric with respect to the vertical axis of the Cartesian plane and are referred to as \textit{even}.
On the other hand, the sine function flips sign when the input does: $\sin(t)=-\sin(-t)$.
Functions with this property are \textit{odd}.
Not all functions are purely even or purely odd.
However, given a real function $f(t)$, it is always possible to compute its even and odd parts
\begin{align}
    f_{\textup{even}}(t)&=\frac{f(t)+f(-t)}{2} \\
    f_{\textup{odd}}(t)&=\frac{f(t)-f(-t)}{2}
\end{align}
such that $f(t)=f_{\textup{even}}(t)+f_{\textup{odd}}(t)$.
The cosine function has zero odd part; the sine function has zero even part; and most functions have a mixture of both.
This also applies to the cross-covariance function.
The Fourier transforms of the even and odd parts of the cross-covariance function produce the real and imaginary parts of the cross-spectral density (\cref{fig:CSD_y12_span} and off-diagonal plots in \cref{fig:CSD_y_0_normalised}).
A special case of the cross-covariance function is the autocovariance function, which is always even.
This is why the diagonal plots in \cref{fig:CSD_y_0_normalised} (the spectral density plots) have zero imaginary part.
We refer the interested reader to \cite{oppenheim1997SignalsSystems} to build further intuition about the Fourier transform and the basics of signal processing.

\section{Analytic solution for the correlation matrix}
\label{sec:correlation_matrix_3by3_analytic}
Integrating the cross-spectral density in \cref{fig:CSD_y_0_normalised} over all frequencies gives us the $3 \times 3$ symmetric correlation matrix $R$, typically used to quantify the functional connectivity.
In this example, we can solve the integral analytically to show the explicit dependence of all the pairwise correlation values on the single effective connectivity parameter $a_{21}$:
\begin{align}
    \rho_{11} &= \rho_{22} = \rho_{33} = 1 \\
    \rho_{12} &= \rho_{21} = \frac{2.7 a_{21}}{\sqrt{17.6 a_{21}^2+10.5}} \\
    \rho_{13} &= \rho_{31} = \frac{5.9 a_{21}-1.4}{\sqrt{a_{21} \left(137 a_{21}-45.7\right)+54.6}} \\
    \rho_{23} &= \rho_{32} = \frac{a_{21} \left(45.7 a_{21}-8.8\right)+13.2}{\sqrt{\left(17.6 a_{21}^2+10.5\right) \left(a_{21} \left(137 a_{21}-45.7\right)+54.6\right)}}.
\end{align}
These functions are plotted in \cref{fig:corr_y}.

\section{DCM diagnostic function and visual report}
\begin{figure}
    \centering\includegraphics[width=0.5\textwidth]{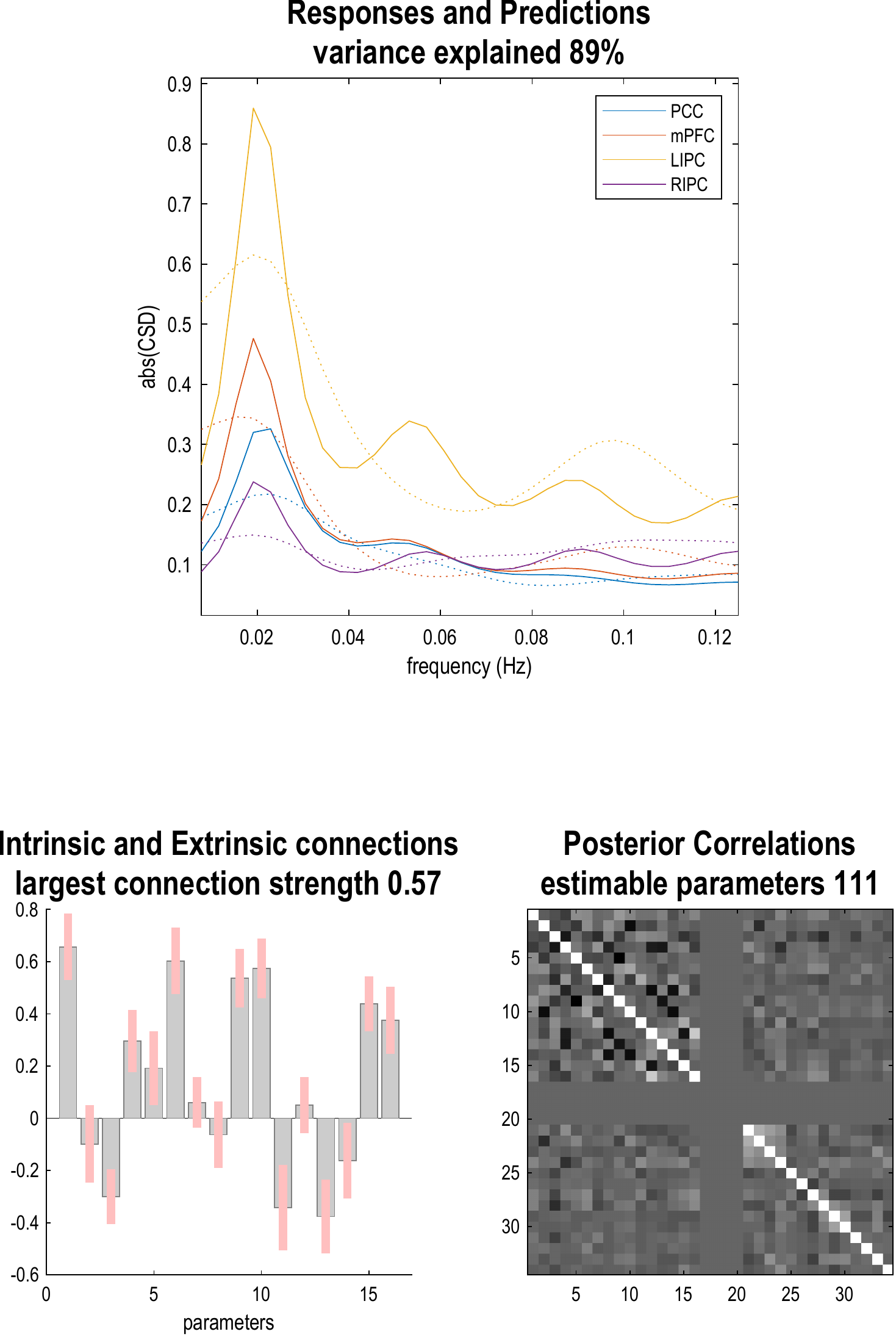}
    \caption{\label{fig:spm_dcm_fmri_check}
        Output of the DCM diagnostics function \texttt{spm\_dcm\_fmri\_check()} generated using SPM12.
        This report indicates the percentage of variance explained ($R^2$), which is a useful performance metric to judge the quality and success of the model fit to the observed cross-spectral density of the data.
        Top: BOLD power spectral density obtained via spectral DCM analysis of a real resting-state fMRI dataset described in \cref{sec:simulated_and_empirical_CSD}.
        The dashed lines represent the predicted power spectral density and the solid lines the observed ones.
        Bottom left: Posterior expectation (grey bars) and posterior variance (pink bars) of the effective connectivity values.
        Bottom right: Posterior correlation matrix between the model parameters.
    }
\end{figure}

\cref{fig:spm_dcm_fmri_check} shows the output of the DCM diagnostics function \texttt{spm\_dcm\_fmri\_check()} generated using SPM12.
This report complements the cross-spectral density plots in \cref{fig:CSD_real_data} and additionally indicates the percentage of variance explained ($R^2$), which is a useful performance metric to judge the quality and success of the model fit to the observed cross-spectral density of the data.
The posterior expectation and variance of the effective connectivity parameters are also plotted, as well as the posterior correlations among all the model parameters.

\end{document}